# Machine learning and natural language processing models to predict the extent of food processing

Nalin Arora[1,3,4], Sumit Bhagat[1,2,3,4], Riya Dhama[1,2,3,4], and Ganesh Bagler[1,3,4,*]

[1]Department of Computational Biology
[2]Department of Computer Science
[3]Infosys Center for Artificial Intelligence
[4]Center of Excellence in Healthcare

Indraprastha Institute of Information Technology Delhi (IIIT-Delhi), New Delhi 110020 India.

**\*Corresponding Author**

Ganesh Bagler (GB): bagler@iiitd.ac.in          ORCID ID: 0000-0003-1924-6070

Infosys Centre for Artificial Intelligence
Center of Excellence in Healthcare
Department of Computational Biology
A-305, R&D Block, Indraprastha Institute of Information Technology Delhi (IIIT-Delhi)
Okhla Phase III,  New Delhi, India, 110020
Phone: 91-11-26907-443

## Abstract

The dramatic increase in consumption of ultra-processed food has been associated with numerous adverse health effects. Given the public health consequences linked to ultra-processed food consumption, it is highly relevant to build computational models to predict the processing of food products. We created a range of machine learning, deep learning, and NLP models to predict the extent of food processing by integrating the FNDDS dataset of food products and their nutrient profiles with their reported NOVA processing level. Starting with the full nutritional panel of 102 features, we further implemented coarse-graining of features to 65 and 13 nutrients by dropping flavonoids and then by considering the 13-nutrient panel of FDA, respectively. LGBM Classifier and Random Forest emerged as the best model for 102 and 65 nutrients, respectively, with an F1-score of 0.9411 and 0.9345 and MCC of 0.8691 and 0.8543. For the 13-nutrient panel, Gradient Boost achieved the best F1-score of 0.9284 and MCC of 0.8425. We also implemented NLP based models, which exhibited state-of-the-art performance. Besides distilling nutrients critical for model performance, we present a user-friendly web server for predicting processing level based on the nutrient panel of a food product: https://cosylab.iiitd.edu.in/food-processing/.

## Acronyms

AUC-ROC - Area Under the Receiver Operating Characteristic Curve; AUPRC - Area Under the Precision-Recall Curve; BERT - Bidirectional Encoder Representations from Transformers; EPIC - European Prospective Investigation into Cancer and Nutrition; ExtraTrees - Extremely Randomized Trees; FDA - Food and Drug Administration; FNDDS - Food and Nutrient Database for Dietary Studies; IARC - International Agency for Research on Cancer; IFIC - The International Food Information Council; IFPRI - International Food Policy Research Institute; KNN - K-Nearest Neighbors; LGBM - Light Gradient Boosting Machine; MCC - Matthews Correlation Coefficient; MLP - Multi-layer Perceptron; NIPH - National Institute of Public Health; NLP - Natural Language Processing; SHAP - SHapley Additive exPlanations; SMOTE - Synthetic Minority Oversampling Technique; SVM - Support Vector Machine; UNC - University of North Carolina at Chapel Hill; UPFs - Ultra-Processed Foods; XGB - Extreme Gradient Boosting

## Highlights

- State-of-the-art ML and DL models to predict the NOVA level.
- State-of-the-art NLP models implementing word embeddings.
- Web server predicting NOVA level based on nutrition concentration.

## 1. Introduction

Food processing is a comprehensive term encompassing all procedures through which raw food ingredients are transformed into consumable, storable, and cookable food products. It spans from fundamental preparations of food to intricate processes, such as altering a food ingredient into another form (*What Is Food Processing? | The Food & Drink Federation*, n.d.). The primary objective of food processing is to extend the shelf life of food items, prevent food spoilage, simplify food storage and transportation, and convert raw food materials into appealing, marketable products. Food processing methods range from basic techniques like chopping, freezing, and heating to more sophisticated methods such as fermentation, emulsification, and the addition of ingredients like preservatives and additives (*What Is Food Processing? | The Food & Drink Federation*, n.d.). Among the most crucial aspects of food processing is food preservation, which involves slowing down the oxidation of fats that could lead to rancidity and inhibiting the growth of fungi, bacteria, and various other microorganisms. It is worth noting, however, that food processing often reduces the nutritional value of food and includes the addition of additives, which can adversely affect the quality of food. Various classification systems have been developed for categorizing food processing methods that offer insights into the diverse approaches used.

At present, various food processing classification systems co-exist, each with its unique approach to categorization. NOVA food processing system, developed in Brazil, is the most widely used of these all (Monteiro et al., n.d.) Among the other systems developed include a classification system developed by The International Food Information Council (IFIC) (Eicher-Miller et al., 2012), and another by the University of North Carolina at Chapel Hill (UNC) (Bleiweiss-Sande et al., 2019; Poti et al., 2015). Furthermore, the International Agency for Research on Cancer (IARC) has formulated a classification system based on methods developed by the European Prospective Investigation into Cancer and Nutrition (EPIC) study (Chajès et al., 2011; Moubarac et al., 2014). Additionally, the National Institute of Public Health in Mexico has constructed a food processing classification system (González-Castell D et al., n.d.; Moubarac et al., 2014) while the International Food Policy Research Institute (IFPRI) has developed one in Guatemala (Asfaw, 2011; Moubarac et al., 2014). For a detailed comparison of the classification systems, refer to Supplementary Data 1, Table S1.

In this study, we have chosen to use the NOVA classification system for its superior quality, as elaborated in the comprehensive review by Moubarc et al. Compared to alternatives like IARC-EPIC, IFIC-JTF, NIPH, and IFPRI, NOVA received the top rating (Moubarac et al., 2014). Known for its precision, coherence, and comprehensiveness, NOVA offers a globally applicable framework and classification system. Its versatility enables easy adaptation across diverse contexts through the incorporation of specific food listings (Moubarac et al., 2014). Moreover, the UNC system, an extension of NOVA, was tailored to categorize products commonly found in US supermarkets (Bleiweiss-Sande et al., 2019; Poti et al., 2015). Widely recognized as a globally accepted approach, NOVA effectively categorizes foods based on their level, scope, and intent of industrial processing. Accordingly, we have chosen to use the NOVA classification system for our study.

Presently, there is a notable surge in the consumption of Ultra-Processed Foods (UPFs), particularly evident in countries like the United States of America and the United Kingdom, where UPFs contribute to over 50% of the total energy intake (Marino et al., 2021). In the United Kingdom, for instance, approximately 65% of middle school children's calorie intake is derived from UPFs (Khandpur et al., 2020). Similarly, in the USA, UPFs account for 66.2% and 66.4% of the total energy intake among school-aged children and adolescents, respectively (Neri et al., 2019). In countries like India, traditional recipes are being transformed into UPFs marketed as convenience foods (Ghosh-Jerath et al., 2023). These trends signify a significant rise in UPF consumption, correlating with the prevalence of diet-related non-communicable diseases (Martínez Steele et al., 2019). Increased UPF intake has also been linked to the rising prevalence of childhood obesity (Coppola et al., 2023) as well as elevated risks of overall and breast cancer, with a 10% increase in UPF proportion in the diet associated with more than a 10% increase in cancer risks (Fiolet et al., 2018). Moreover, heightened UPF consumption is associated with increased risks of cardiovascular, coronary heart, and cerebrovascular diseases (Srour et al., 2019). Notably, elevated UPF consumption impacts not only physical health but also mental well-being, with studies indicating a positive association between UPF consumption and incident depressive symptoms (Adjibade et al., 2019). Research conducted in South Korea demonstrated a direct correlation between UPF consumption and depression among Korean women, while a study in Melbourne, Australia, found that higher intake of ultra-processed foods was linked to subsequent elevated psychological distress, indicative of depression (Lane et al., 2023; Lee & Choi, 2023). Given the concerning health implications of UPFs, there is a growing recognition for the need for effective classification systems to better understand and address this dietary challenge. Consequently, exploring innovative approaches like developing an automated classifier based on the NOVA system is imperative to enhance our understanding and management of UPF consumption and its associated health risks. Analysis of Figure 1 suggests a noticeable shift in nutrient concentration across NOVA classes, highlighting the potential efficacy of employing a data-driven approach (For more examples, refer to Figure S1 and Figure S2 in Supplementary Data 1). The current NOVA system, being labor-intensive and reliant on manual classification, necessitates automation to streamline the process and improve efficiency (Bleiweiss-Sande et al., 2019; Braesco et al., 2022). We hypothesize that processing food alters its nutritional profile, and therefore, it is possible to build good quality machine learning models for predicting food processing level by using nutritional profiles as a feature.

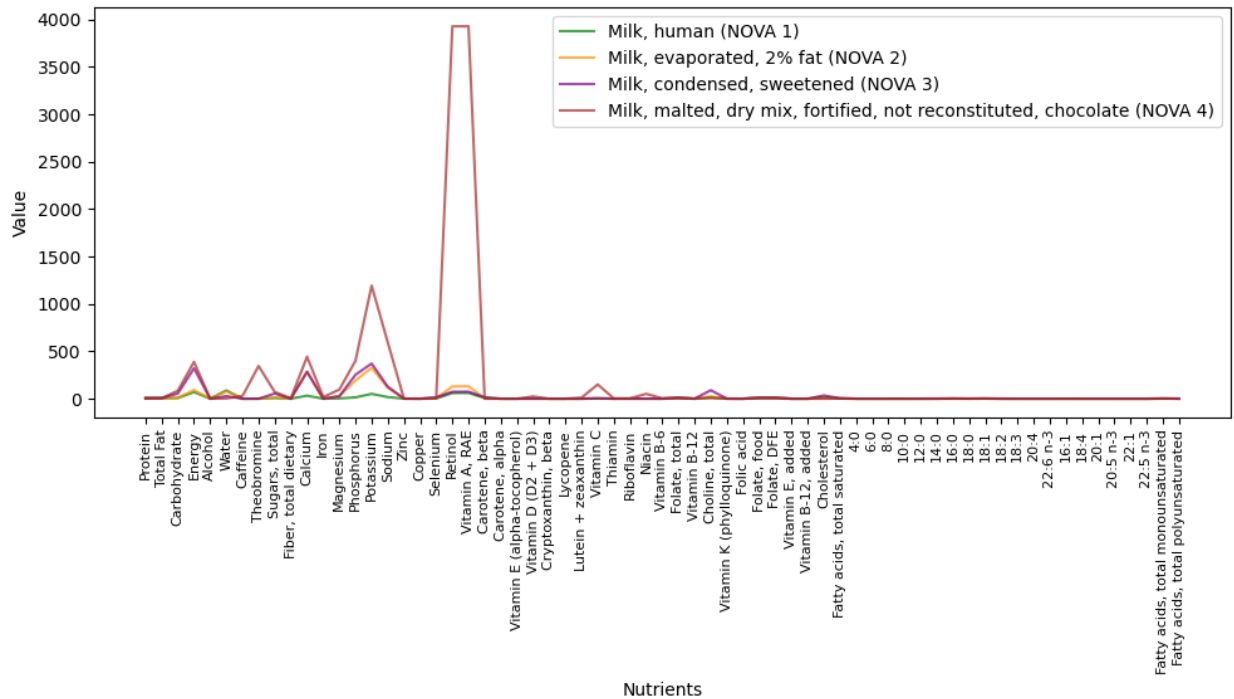

**Figure 1: Nutrient concentration comparison for milk products with varying NOVA classes**

Among the previous attempts to automate food processing labels, some have utilized machine learning and large language models for food classification purposes. Menichetti et al. developed a random forest-based algorithm for classifying foods into NOVA categories (Menichetti et al., 2023a). Another study by Ignacio Fernández-Villacañas Marcos et al. employed various unsupervised clustering learning techniques based on K-means, hierarchical, probabilistic, and spectral clustering to classify food products (Ignacio Fernández-Villacañas Marcos et al., 2022). Most recently, Hu et al. utilized large language models to classify food products into NOVA (Hu et al., 2023).

The study conducted by Menichetti et al. offers valuable insights through their focus on the Random Forest model and the use of AUC-ROC (Area Under the Receiver Operating Characteristic Curve) and AUPRC (Area Under the Precision Recall Curve) scores (Menichetti et al., 2023a). However, their approach primarily addresses the performance metrics on just Random Forest, which might not fully capture the complexities involved in multi-class classification scenarios, especially when classifying ultra-processed foods (UPFs). AUC-ROC and AUPRC, while commonly used, have limitations in contexts with class imbalances. These metrics do not provide nuanced insights into how well the model distinguishes between different classes or the performance of minority classes. Moreover, the work overlooks an essential nutrient: the energy content in food products. As highlighted by Gupta et al., energy content plays a crucial factor in analyzing food patterns and nutritional profiles(Gupta et al., 2019a). Incorporating energy content into nutritional analysis can provide deeper insights into food patterns and dietary health impacts, revealing the true nutritional quality of food products.

Building on these insights, our work adopts a more holistic approach to predict the food processing level of food products. We employ a diverse range of machine learning techniques, including 11 different machine learning models and one deep learning model, to provide a broader perspective on classification performance. Our evaluation extends beyond AUC-ROC and AUPRC scores to include metrics such as Accuracy, Precision, Recall, F1-Score, and Matthews Correlation Coefficient (MCC), ensuring a more thorough assessment of model efficacy. To address class imbalance, we implement a refined strategy combining SMOTE (Synthetic Minority Oversampling Technique) with stratified k-fold cross-validation, enhancing the reliability of our results as well as prevent overfitting. Furthermore, we incorporate pre-trained word embeddings to enrich feature representation, capturing semantic relationships within food descriptions, categories, ingredients, and the nutrient panels. Importantly, our study integrates the energy content of food products into the nutrient panel, recognizing its role in assessing nutritional value and health outcomes. By considering energy content, we aim to improve the depth of nutritional analysis while improving the accuracy of food processing classification. Such a rigorous methodology is expected to improve the model's performance and provide deeper insights. By leveraging multiple Machine Learning models, NLP-based models and a wide range of evaluation metrics, we ensure the robustness and reliability of our findings.

## 2. Materials and methods

### 2.1 Dataset description

The nutrient data utilized in this study was sourced from the Food and Nutrient Database for Dietary Studies (FNDDS) 2009-2010 (*FNDDS DOWNLOAD DATABASES : USDA ARS*, n.d.). FNDDS is a comprehensive database maintained by the U.S. Department of Agriculture (USDA), providing detailed information on the nutritional content of various foods commonly consumed in the United States. To categorize foods based on their processing levels, we employed NOVA labels obtained from Menichetti et al. (Menichetti et al., 2023a) and Steele et al (Steele et al., 2016). To begin with, our nutrient panel included all 102 nutrient parameters (Table S2 in Supplementary Data 1), including 37 flavonoids extracted from the FNDDS data spanning 2007-2010. For subsequent studies, we focused on 65 nutrients to remove the 37 flavonoids' data (Table S3 in Supplementary Data 1). In the final phase, we considered 13 specific nutrients mandated by the FDA (*The Nutrition Facts Label | FDA*, n.d.) (Table S4 in Supplementary Data 1). In total, we analyzed 2970 food items, each associated with a NOVA classification label. This comprehensive dataset from FNDDS facilitated a thorough exploration of the relationships between nutrient profiles and food processing levels across a diverse range of food items.

### 2.2 Data pre-processing

We extensively preprocessed to prepare the dataset for subsequent analysis. Firstly, the FNDDS data was integrated with their NOVA labels obtained from Menchetti et al and Steele et al (Menichetti et al., 2023a; Steele et al., 2016). This integration was pivotal for aligning the food items with their respective NOVA classifications. Subsequently, nutrient codes within the FNDDS data were replaced with their corresponding nutrient names. This transformation enhanced the interpretability of the dataset, facilitating a nuanced understanding of the nutritional content of various food items.

### 2.3 Model development

A variety of machine learning algorithms were considered to predict the NOVA food processing level of food products, starting from baseline models and moving to more complex ensemble methods. A total of 12 algorithms were evaluated, utilizing the scikit-learn library for most models, except for Light Gradient Boost Machine (LGBM) and Extreme Gradient Boosting (XGB), which were implemented using their respective libraries. The following algorithms were included in the study: Logistic Regression, Support Vector Machines (SVM), Decision Trees, K-Nearest Neighbors (KNN), Naive Bayes, Random Forest, XGB, Gradient Boost, AdaBoost, Extremely Randomized Trees (ExtraTrees), LGBM Classifier, Multi-Layer Perceptron.

Baseline algorithms such as Logistic Regression, a classification algorithm that models the probability of a binary outcome (Maalouf, 2011), SVM, a supervised learning algorithm that finds the optimal hyperplane to separate different classes (Evgeniou & Pontil, 2001) , Decision Trees, a model that splits data points into branches to make predictions based on the features (Song & Lu, 2150), KNN is a non-parametric method that classifies using the majority class among the k-nearest neighbors (Cunningham & Delany, 2021), and Naive Bayes, classifier based on Bayes' theorem assuming feature independence (Yang, 2018), were initially employed, followed by the introduction of more sophisticated ensemble techniques. Random Forest, an ensemble method that constructs multiple decision trees and merges them to improve performance (Breiman, 2001), XGB, an optimized gradient boosting framework designed to be highly efficient (Chen & Guestrin, 2016), Gradient Boost, an ensemble model that builds trees sequentially, each tree correcting the errors made by the previous trees (He et al., 2019), AdaBoost, an ensemble method that combines weak classifiers to create a strong classifier by improving on errors of previous models (Tu et al., 2017), and ExtraTrees, an ensemble method that builds multiple trees with random splits for each feature (Geurts et al., 2006), were then introduced to capture more complex relationships within the data. Finally, LGBM, a highly efficient gradient boosting framework that uses tree-based algorithms (Ke et al., n.d.), and MLP , a type of neural network that consists of multiple layers of nodes to learn complex relationships in the data (Balas et al., 2009), models were incorporated to further enhance predictive performance. RandomizedSearchCV from scikit-learn, a hyperparameter optimization technique that performs a random search over specified parameter values for a model to enhance performance (Bergstra et al., 2012), was used to hyperparameter-tune the parameters of the models to further enhance the performance of the models.

**2.4 Handling class imbalance**

Given the significant class imbalance in the dataset, three techniques were employed to mitigate its effects: SMOTE, which creates synthetic samples of the minority class to balance the dataset using the previous entries (Chawla et al., 2002), stratified k-fold cross-validation, which ensures that each fold of cross-validation has the same proportion of classes as the original dataset (Prusty et al., 2022), and a combination of SMOTE and stratified k-fold, which applies SMOTE within each fold of stratified cross-validation to maintain balance throughout the training process.

**2.5 Evaluation metrics**

A diverse set of evaluation metrics was employed to provide a holistic assessment model performance. The following metrics were utilized: Accuracy, F1-Score, Precision, Recall, Matthews Correlation Coefficient (MCC), and Receiver Operating Characteristic (ROC) curves. These evaluation metrics offer comprehensive insights into the predictive performance of each model, encompassing various aspects of classification accuracy, balance, and robustness.

**2.6 Feature importance analysis**

To gain insights into the factors driving the predictions of the best-performing models, SHAP (SHapley Additive exPlanations) analysis was conducted. SHAP is a game-theoretic approach to explain the output of any machine learning model. It provides a unified measure of feature importance by quantifying the contribution of each feature to the model's predictions (Lundberg & Lee, 2017). For the best-performing models identified during the evaluation phase, SHAP analysis was applied to uncover the relative importance of features in predicting the NOVA food processing level.

**2.7 Pre-trained word embeddings**

We augmented our dataset with three additional columns: 'Food Description', 'Category Name', and 'Macro Class.' To transform these textual features into numerical representations suitable for machine learning algorithms, we utilized several pre-trained embeddings, including Bidirectional Encoder Representations from Transformers (BERT)(Devlin et al., 2018) and its variants: LegalBERT, DistilBERT, and XLM-RoBERTa, along with GPT-2. BERT and its variants work exceedingly well to capture the bidirectional context in text data, and they are encoder-only models, which enables them to generate rich contextual embeddings. LegalBERT is fine-tuned explicitly for legal text (Chalkidis et al., n.d.), DistilBERT is a refined version of BERT that offers faster inference with a smaller architecture (Sanh et al., 2019), and XLM-RoBERTa is designed for multilingual text (Conneau et al., n.d.). GPT-2, on the other hand, is a transformer-based language model, a decoder-only model, developed by OpenAI, generating contextual embeddings by predicting the next word in a sequence based on preceding words (Tan et al., 2020). These embeddings encode semantic information inherent in the food descriptions, category names, and macro class labels, enriching the feature space alongside the existing nutrient panels for improved model performance. Subsequently, various machine learning algorithms were applied to the combined feature set, leveraging both nutrient panels and derived embeddings to enhance predictive capability in nutrient prediction tasks.

## 3. Results

**3.1 Data characterization**

We observe that the data is severely imbalanced, with over-representation of NOVA Class 4 food products and under-representation of Class 2 (Supplementary Data 1, Figure S3). Additionally, t-SNE reduced dataset dimensionality, revealing patterns in a lower-dimensional space. Leveraging t-SNE aimed to uncover underlying structures for further analysis (Van Der Maaten & Hinton, 2008). Supplementary Data 1, Figure S4 illustrates the separability achieved using t-SNE with a perplexity of 50, which led to data points converging with their groups, indicating a trend towards a unified representation of underlying structures. This observation highlights t-SNE's effectiveness in revealing patterns within high-dimensional data, offering nuanced insights into the interrelation between nutrient profiles and NOVA classification labels. This finding contributes to a better understanding of how processing levels manifest in the nutrient space, guiding future analyses and dietary considerations.

Following this, we conducted a meticulous check for missing data to ensure the integrity and completeness of the dataset. This step involved a thorough examination of all variables to identify any instances of missing or incomplete information. Moreover, to enhance the comparability and interpretability of the features, the data was subjected to standardization using the widely adopted scikit-learn library's StandardScaler (Conneau et al., n.d.). This step involved rescaling the features to have a mean of zero and a standard deviation of one, a critical procedure in many machine learning applications. By standardizing the data, we ensured that all variables contributed equally to the subsequent analyses, preventing any undue influence from features with larger scales.

| Model | F1 Score | MCC | Accuracy | Precision | Recall |
|---|---|---|---|---|---|
| LGBM | 0.9411 | 0.8691 | 0.9411 | 0.9419 | 0.9411 |
| Gradient Boost | 0.9307 | 0.8463 | 0.931 | 0.9312 | 0.931 |
| Random Forest | 0.9299 | 0.8455 | 0.9293 | 0.9308 | 0.9293 |
| XGB | 0.9212 | 0.8294 | 0.9192 | 0.9254 | 0.9192 |
| ExtraTrees | 0.9106 | 0.8017 | 0.9108 | 0.9112 | 0.9108 |
| MLP | 0.9028 | 0.7856 | 0.9024 | 0.9042 | 0.9024 |
| KNN | 0.8762 | 0.7304 | 0.8721 | 0.8834 | 0.8721 |
| Decision Trees | 0.8665 | 0.7074 | 0.8636 | 0.8726 | 0.8636 |
| Logistic Regression | 0.8093 | 0.6274 | 0.7946 | 0.8484 | 0.7946 |
| SVM | 0.7743 | 0.5802 | 0.7508 | 0.8409 | 0.7508 |
| AdaBoost | 0.589 | 0.3397 | 0.5522 | 0.7425 | 0.5522 |
| Naïve Bayes | 0.5589 | 0.31 | 0.4882 | 0.792 | 0.4882 |

**Table 1: Model Performance for the 102 nutrients panel.**

### 3.2 Models with 102 nutrient features

To begin with, we built models by accounting for all the 102 features. Table 1 presents the performance metrics of various machine learning models trained on a dataset featuring a 102-nutrient panel (Supplementary Data 2, Table S1). The ensemble model LGBM (trained with SMOTE and k-fold) was presented with a state-of-the-art performance, achieving an impressive F1-score of 0.9411 and an MCC score of 0.8691. Additionally, Gradient Boosting, Random Forest, XGB, and ExtraTrees also exhibit good performances, with F1-scores of 0.9307, 0.9299, 0.9212, and 0.9106 respectively, accompanied by MCC scores of 0.8463, 0.8455, 0.8294, and 0.8017 respectively, showcasing the efficacy of ensemble methods in learning complex patterns within the 102-nutrient data for accurate predictions. While non-ensemble models like MLP and KNN demonstrate respectable performance with F1-scores of 0.9028 and 0.8762, respectively, and MCC scores of 0.7856 and 0.7304, respectively, they fall short compared to ensemble approaches. Decision Trees, Logistic Regression, SVM, and AdaBoost, on the other hand, exhibit relatively weaker performances, with F1-scores of 0.8665, 0.8093, 0.7743, and 0.5890 respectively, and MCC scores of 0.7074, 0.6274, 0.5802 and 0.3397 respectively, indicating their limitations in capturing the complexities inherent in the nutrient panel dataset. Naive Bayes returned the lowest performance metrics, with an F1-score of 0.5589 and an MCC of 0.3100, indicating its inadequacy in handling the complexity of the dataset. Refer to Supplementary Data 1, Figure S5 for the best ROC and precision-recall curves. Refer to Supplementary Data 2, Table S2 for the model performance using SMOTE and stratified k-fold, and Table S3 for only stratified k-fold.

| Model | F1 Score | MCC | Accuracy | Precision | Recall |
|---|---|---|---|---|---|
| **Random Forest** | 0.9388 | 0.8648 | 0.9377 | 0.941 | 0.9377 |
| **LGBM** | 0.9309 | 0.8473 | 0.9293 | 0.9338 | 0.9293 |
| **XGB** | 0.9291 | 0.8444 | 0.9276 | 0.9318 | 0.9276 |
| **Gradient Boost** | 0.9238 | 0.832 | 0.9226 | 0.9262 | 0.9226 |
| **ExtraTrees** | 0.9131 | 0.8075 | 0.9125 | 0.9146 | 0.9125 |
| **MLP** | 0.8844 | 0.7454 | 0.8805 | 0.8911 | 0.8805 |
| **KNN** | 0.8639 | 0.7065 | 0.8586 | 0.8738 | 0.8586 |
| **Decision Trees** | 0.8522 | 0.6809 | 0.8468 | 0.8632 | 0.8468 |
| **Logistic Regression** | 0.7936 | 0.6143 | 0.7761 | 0.8478 | 0.7761 |
| **SVM** | 0.78 | 0.593 | 0.7576 | 0.8468 | 0.7576 |
| **AdaBoost** | 0.5145 | 0.3957 | 0.4781 | 0.7576 | 0.4781 |
| **Naïve Bayes** | 0.524 | 0.2904 | 0.4579 | 0.7761 | 0.4579 |

**Table 2: Model Performance for the 65 nutrients panel.**

### 3.3 Models with 65 nutrient features

Going further, we used a coarse-grained set of features with only 65 nutrients by removing the flavonoids. Table 2 presents the performance metrics of various machine learning models trained on a dataset featuring the 65-nutrient panel (Supplementary Data 3, Table S2). The ensemble model Random Forest (trained with SMOTE and stratified k-fold) demonstrates state-of-the-art performance, achieving an impressive F1-score of 0.9388 and an MCC score of 0.8646. Despite a drastic reduction in the features, the best model's performance was only marginally compared to that obtained from a panel of 102 nutrients. Additionally, LGBM, XGB, Gradient Boost, and ExtraTrees also exhibit good performances, with F1-scores of 0.9309, 0.9291, 0.9238, and 0.9131 respectively, accompanied by MCC scores of 0.8473, 0.8444, 0.8320, and 0.8075 respectively,

showcasing the efficacy of ensemble methods in learning complex patterns within the 65-nutrient data for accurate predictions. While non-ensemble models like MLP and KNN demonstrate respectable performance with F1-scores of 0.8844 and 0.8639, respectively, and MCC scores of 0.7454 and 0.7065, respectively, they fall short compared to ensemble approaches. Decision Trees, Logistic Regression, SVM, and AdaBoost, on the other hand, exhibit relatively weaker performances, with F1-scores of 0.8522, 0.7936, 0.7800, and 0.5145 respectively, and MCC scores of 0.6809, 0.6143, 0.5930 and 0.3057 respectively, indicating their limitations in capturing the complexities inherent in the nutrient panel dataset. Naive Bayes stands out with the lowest performance metrics, with an F1-score of 0.5240 and an MCC of 0.2904, indicating its inadequacy in handling the complexity of the dataset. Refer to Supplementary Data 1, Figure S6 for the best ROC and precision-recall curves. Refer to Supplementary Data 3, Table S1 for the model performance using only SMOTE, and Table S3 for only stratified k-fold.

| Model | F1 Score | MCC | Accuracy | Precision | Recall |
|---|---|---|---|---|---|
| **Gradient Boost** | 0.9284 | 0.8425 | 0.9276 | 0.9297 | 0.9276 |
| **XGB** | 0.9209 | 0.8286 | 0.9192 | 0.9246 | 0.9192 |
| **LGBM** | 0.919 | 0.8225 | 0.9175 | 0.9216 | 0.9175 |
| **Random Forest** | 0.9206 | 0.8224 | 0.9192 | 0.9232 | 0.9192 |
| **ExtraTrees** | 0.92 | 0.8216 | 0.9192 | 0.9228 | 0.9192 |
| **KNN** | 0.8882 | 0.7545 | 0.8855 | 0.8941 | 0.8855 |
| **MLP** | 0.8733 | 0.7194 | 0.8721 | 0.8786 | 0.8721 |
| **Decision Trees** | 0.8549 | 0.6885 | 0.8485 | 0.8663 | 0.8485 |
| **SVM** | 0.7815 | 0.5861 | 0.7626 | 0.8352 | 0.7626 |
| **AdaBoost** | 0.7383 | 0.5251 | 0.7155 | 0.8107 | 0.7155 |
| **Logistic Regression** | 0.6573 | 0.4283 | 0.6212 | 0.7869 | 0.6212 |
| **Naïve Bayes** | 0.4784 | 0.2752 | 0.4411 | 0.7388 | 0.4411 |

**Table 3: Model Performance for the 13 nutrients panel prescribed by FDA.**

### 3.4 Models with 13 nutrients features

Coarse-graining further down, we created a panel of 13 nutrients mandated by the FDA (*The Nutrition Facts Label | FDA*, n.d.). Table 3 presents the performance metrics of various machine learning models trained on a dataset featuring a 13-nutrient panel (Supplementary Data 4, Table S2). The ensemble model Gradient Boost (trained with SMOTE and stratified k-fold) demonstrates state-of-the-art performance, achieving an impressive F1-score of 0.9284 and an MCC score of 0.8425. While the scores dropped from the full panel model trained with 102 and 65 parameters, the best model still presents an impressive performance with limited training data. Additionally, XGB, LGBM, Random Forest, and ExtraTrees also exhibit strong performances, with F1-scores of 0.9209, 0.9190, 0.9206, and 0.9200, respectively, accompanied by MCC scores of 0.8286,

0.8225, 0.8224, and 0.8216 respectively, showcasing the efficacy of ensemble methods in learning complex patterns within the 13-nutrient data for accurate predictions. While non-ensemble models like KNN and MLP demonstrate respectable performance with F1-scores of 0.8882 and 0.8733, respectively, and MCC scores of 0.7545 and 0.7194, respectively, they fall short compared to ensemble approaches. Decision Trees, SVM, AdaBoost, and Logistic Regression, on the other hand, exhibit relatively weaker performances, with F1-scores of 0.8549, 0.7815, 0.7383, and 0.6573 respectively, and MCC scores of 0.6885, 0.5861, 0.5251 and 0.4283 respectively, indicating their limitations in capturing the complexities inherent in the nutrient panel dataset. Naive Bayes presented the lowest performance metrics, with an F1-score of 0.4784 and MCC of 0.2752, indicating its inadequacy in handling the complexity of the dataset. Refer to Supplementary Data 1, Figure S7 for the best ROC and precision-recall curves. Refer to Supplementary Data 4, Table S1 for the model performance using only SMOTE, and Table S3 for only stratified k-fold.

### 3.5 Comparison of model performance with previous studies

We performed a one-to-one comparison of our results with the previous state-of-the-art from Menichetti et al (Menichetti et al., 2023a). Menichetti et al. utilized nutrient panels of 12, 62, and 99 nutrients for their respective analyses by leaving three nutrients, namely, 'Energy', 'Folate, DFE', and 'Vitamin A, RAE.' We experimented with the same set of nutrients for 12, 62, and 99 nutrients. Their FoodProX algorithm based on Random Forest Classifier achieved AUC scores of 0.98, 0.97, 0.97, and 0.98, and AUP scores of 0.89, 0.76, 0.86, and 0.99, respectively for NOVA 1, NOVA 2, NOVA 3 and NOVA 4 for the 12-nutrient dataset. In our experiment, we found Random Forest Classifier surpassing these results with AUC scores of 0.98, 0.99, 0.97, and 0.98 and AUP scores of 0.90, 0.78, 0.86, and 0.99, respectively, for NOVA 1, NOVA 2, NOVA 3 and NOVA 4 (Supplementary Data 1, Figure S8 for the ROC and precision-recall curves). For the 62-nutrient dataset, Menichetti et al. achieved AUC scores of 0.98, 0.96, 0.97, and 0.98, and AUP scores of 0.89, 0.73, 0.87, and 0.99, respectively, for NOVA 1, NOVA 2, NOVA 3 and NOVA 4. In contrast, LGBM trained during our experiment yielded higher AUC scores of 0.99, 0.98, 0.98, and 0.98, and AUP scores of 0.91, 0.78, 0.90, and 0.99, respectively, for NOVA categories 1, 2, 3, and 4 (Supplementary Data 1, Figure S9 for the ROC and precision-recall curves). Similarly, for the 99-nutrient dataset, Menichetti et al. reported AUC scores of 0.98, 0.96, 0.97, and 0.98, and AUP scores of 0.88, 0.75, 0.88, and 0.99. In contrast, LGBM trained during our study yielded higher AUC scores of 0.99, 0.98, 0.98, and 0.98, and AUP scores of 0.92, 0.80, 0.91, and 0.99, respectively, for NOVA categories 1, 2, 3, and 4 (Supplementary Data 1, Figure S10 for the ROC and precision-recall curves). These findings highlight a better predictive performance by the models developed in the present study than those reported by Menichetti et al. Our models consistently yielded higher AUC and AUP scores across various nutrient datasets and NOVA categories.

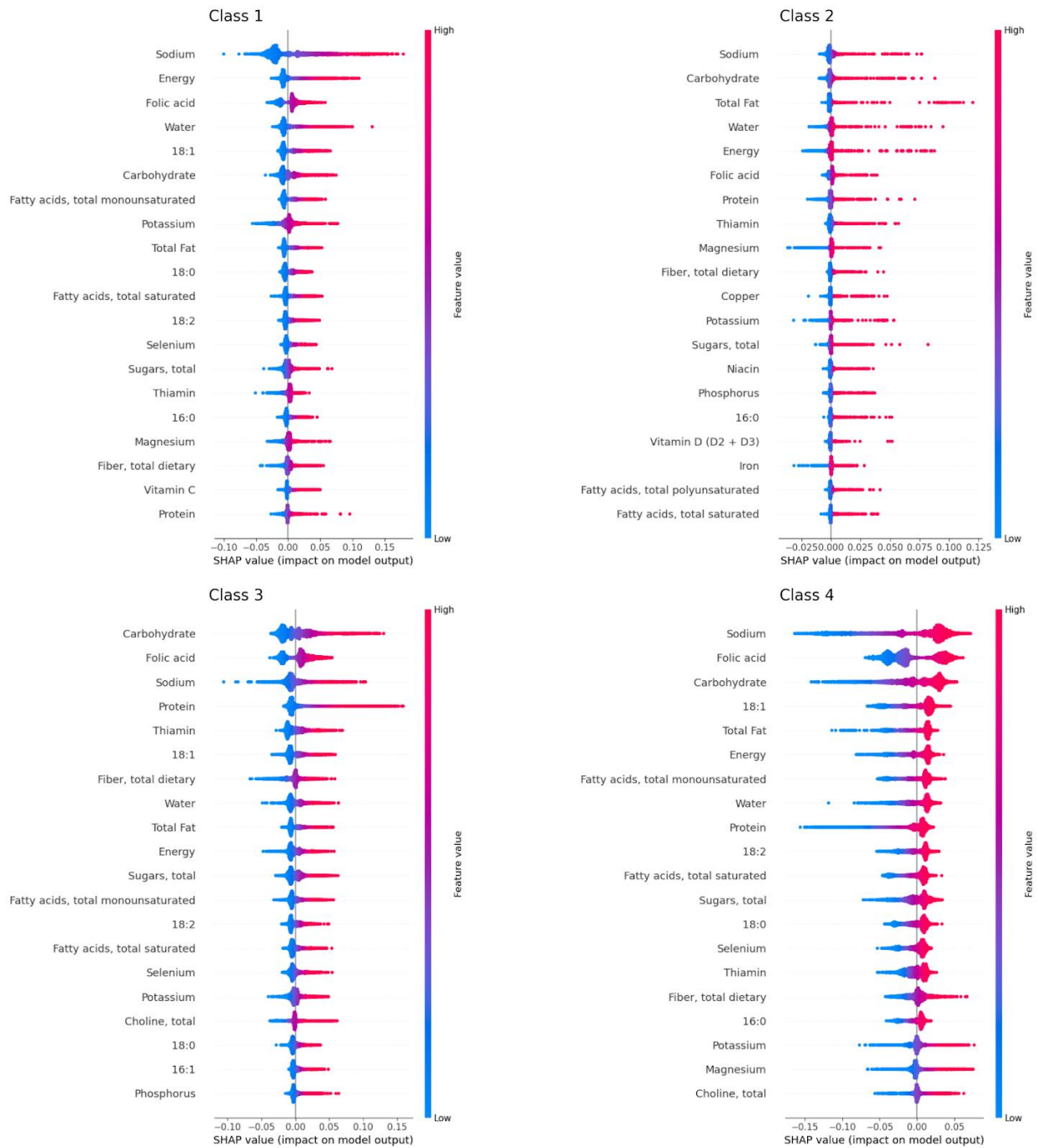

**Figure 2: SHAP summary plots for 102 nutrients panel.**

### 3.6 Feature importance

The SHAP analysis facilitated the identification of key nutrients and attributes that significantly influence the classification outcomes. By leveraging SHAP, we aimed to enhance interpretability and understanding of the underlying mechanisms that drive the predictive performance of the models. For NOVA Class 1, we infer that the top 5 important features are 'Sodium', 'Energy',

'Folic Acid', 'Water', and '18:1'(oleic acid) (Figure 2). For NOVA Class 2, we conclude that the top 5 important features are 'Sodium', 'Carbohydrate', 'Total Fat', 'Water' and 'Energy'. For NOVA Class 3, we identified 'Carbohydrate', 'Folic Acid', 'Sodium', 'Protein', and 'Thiamin' as the top 5 important features (from Figure 2). 'Sodium', 'Folic Acid', 'Carbohydrate', '18:1'(oleic acid), and 'Total Fat' emerged as the top 5 features for NOVA class 4. The insights gained from feature importance analysis contribute to a deeper understanding of the relationship between nutrient profiles and food processing levels, informing future research and dietary considerations. Refer to Supplementary Data 1, Figure S11 and Figure S12 for SHAP summary plots for 65 nutrients and 13 nutrients, respectively.

### 3.7 Pre-trained word-embeddings

Further, we enriched the features by integrating pre-trained word embeddings for 'Food Description', 'Category Name', and 'Macro Class.' These NLP-based features are expected to capture relevant food processing subtleties to enhance the model's performance.

| Model | F1 Score | MCC | Accuracy | Precision | Recall |
|---|---|---|---|---|---|
| Gradient Boost | 0.9528 | 0.8956 | 0.9529 | 0.9537 | 0.9529 |
| LGBM | 0.9482 | 0.8861 | 0.9478 | 0.9494 | 0.9478 |
| XGB | 0.9439 | 0.8775 | 0.9426 | 0.9462 | 0.9428 |
| Random Forest | 0.938 | 0.8632 | 0.9377 | 0.9385 | 0.9377 |
| ExtraTrees | 0.8503 | 0.6972 | 0.8367 | 0.8851 | 0.8367 |

**Table 4: Model performance for the best pre-trained word embedding (XLM-RoBERTa).**

### 3.7.1 102 nutrients panel

Table 4 presents the performance metrics of various machine learning models trained on a dataset featuring a 102-nutrient panel and the textual columns converted to numerical vectors using XLM-RoBERTa (Supplementary Data 5, Table S5). Gradient Boost achieves the highest predictive capability with an F1-score of 0.9528 and an MCC score of 0.8956. Gradient Boost is closely followed by LGBM, XGB, and Random Forest, demonstrating an F1-score of 0.9482, 0.9439, and 0.9380, respectively, and an MCC score of 0.8861, 0.8775, and 0.8632 respectively. ExtraTrees gives a respectable performance, albeit significantly lower than Gradient Boost, with an F1-score of 0.8503, respectively, and an MCC score of 0.6972. Refer to Supplementary Data 1, Figure S21 for the best ROC and precision-recall curves. Refer to Supplementary Data 5, Table S1, Table S2, Table S3, and Table S4, respectively, for model performance using BERT, DistilBERT, GPT-2, and LegalBERT pre-trained word embeddings.

| Model | F1 Score | MCC | Accuracy | Precision | Recall |
|---|---|---|---|---|---|
| LGBM | 0.9518 | 0.8939 | 0.9512 | 0.9529 | 0.9512 |
| Gradient Boost | 0.9414 | 0.8708 | 0.9411 | 0.9418 | 0.9411 |
| Random Forest | 0.938 | 0.8623 | 0.9377 | 0.9385 | 0.9377 |
| XGB | 0.9225 | 0.842 | 0.9226 | 0.933 | 0.9226 |
| ExtraTrees | 0.8629 | 0.7254 | 0.519 | 0.8941 | 0.8519 |

**Table 5: Model performance for the best pre-trained word embedding (DistilBERT).**

### 3.7.2 65 nutrients panel

Table 5 presents the performance metrics of various machine learning models trained on a dataset featuring a 65-nutrient panel and the textual columns converted to numerical vectors using DistilBERT (Supplementary Data 6, Table S2). LGBM achieves the highest predictive capability with an F1-score of 0.9518 and an MCC score of 0.8939. LGBM is closely followed by Gradient Boost, Random Forest, and XGB, yielding an F1-score of 0.9414, 0.9380, and 0.9255, respectively, and an MCC score of 0.8708, 0.8623, and 0.8420 respectively. ExtraTrees also gives a respectable performance, albeit significantly lower than LGBM, with an F1-score of 0.8629, respectively, and an MCC score of 0.7254. Refer to Supplementary Data 1, Figure S22 for the best ROC and precision-recall curves. Refer to Supplementary Data 6, Table S1, Table S3, Table S4, and Table S5, respectively, for model performance using BERT, GPT-2, LegalBERT, and XLM-RoBERTa pre-trained word embeddings.

| Model | F1 Score | MCC | Accuracy | Precision | Recall |
|---|---|---|---|---|---|
| LGBM | 0.9583 | 0.9091 | 0.9579 | 0.9593 | 0.9579 |
| Gradient Boost | 0.9567 | 0.9052 | 0.9562 | 0.9577 | 0.9562 |
| XGB | 0.955 | 0.9018 | 0.9545 | 0.956 | 0.9545 |
| Random Forest | 0.9383 | 0.8646 | 0.9377 | 0.9394 | 0.9377 |
| ExtraTrees | 0.8823 | 0.76 | 0.8737 | 0.906 | 0.8737 |

**Table 6: Model performance for the best pre-trained word embedding (GPT-2).**

### 3.7.3 13 nutrients panel

Table 6 presents the performance metrics of various machine learning models trained on a dataset featuring a 13-nutrient panel and the textual columns converted to numerical vectors using GPT-2 (Supplementary Data 7, Table S3). LGBM achieves the highest predictive capability with an F1-score of 0.9583 and an MCC score of 0.9091. LGBM is closely followed by Gradient Boost, XGB, and Random Forest, yielding F1-score of 0.9567, 0.9550, and 0.9383, respectively, and MCC scores of 0.9052, 0.9018, and 0.8646 respectively. ExtraTrees also gives a respectable performance, albeit significantly lower than LGBM, with F1-score of 0.8823, respectively, and MCC score of 0.7600. Refer to Supplementary Data 1, Figure S23 for the best ROC and precision-recall curves. Refer to Supplementary Data 7, Table S1, Table S2, Table S4, and Table S5, respectively for model performance using BERT, DistilBERT, LegalBERT, and XLM-RoBERTa pre-trained word embeddings.

## 4. Discussion

In summary, this article tackles the problem of data-driven prediction of food processing labels using nutrient panels as a feature. Other than superseding a previous model in performance, we have implemented a range of machine-learning and deep-learning models to compare their performances. Besides implementing SHAP analysis to identify features central to the predictive ability of the model, we have added a new layer of NLP-based features that yielded state-of-the-art performance. We present a web-server based on the best model to predict NOVA food processing labels. We, therefore, surmise that the hypothesis linking the nutritional profile of ingredients with their food processing level and the feasibility of building statistical models for prediction of food processing level stands vindicated.

Notwithstanding the rigor with which the studies were implemented, the dataset used to train the models is relatively small, comprising only 2970 entries. This constraint may affect the model's ability to learn robust patterns and generalize effectively to diverse dietary contexts. While the selected nutrients are universally recognized, reliance solely on the FNDDS data, which primarily represents dietary patterns of the USA, challenges the extension of the model's applicability globally. Variations in dietary habits across regions and cultures may need to be adequately captured, potentially expanding the model's accuracy outside the North American diet. The dataset utilized for model training was limited to the years 2009-2010. This could affect the model's relevance to current dietary trends, technological advances, and economic shifts. Failure to account for these temporal variations may impact the model's ability to adapt to evolving dietary patterns.

One of the issues with the present dataset is class imbalance. To evaluate the impact of class imbalance, primarily due to small size of NOVA Class 2 on the efficacy of models, we conducted experiments of (a) merger of NOVA Class 2 with Class 3, and (b) omission of NOVA Class 2 altogether. We found that neither the merger nor the omission affects the classification ability of the models significantly (Section 10 of Supplementary Data 1).

Going forward, the dataset needs to be augmented by integrating data from diverse sources to account for present-day dietary patterns from across the globe. Such a dataset will help improve the model's training and improve its generalizability with a wider variety of dietary patterns and nutritional compositions. Besides extending the dataset, it is crucial to ensure that the data includes corresponding NOVA labels or incorporates another classification system for food processing. This step is essential for achieving a more nuanced understanding of how food processing impacts nutritional profiles. By selecting datasets with comprehensive labeling or alternative classification methods, we can gather better insights that will aid in developing more precise predictive models. We have demonstrated the utility of NLP techniques in improving food processing predictions. NLP can effectively analyze textual data associated with food descriptions, labels, and ingredients, thereby enhancing the accuracy and granularity of predictions. More rigorous integration of NLP techniques can significantly enhance the model performance.

## 5. Conclusion

This research effectively addresses the prediction of NOVA food processing level using nutrient concentration, surpassing previous models in performance. By employing a range of machine-learning and deep-learning models and integrating NLP-based features, we achieved state-of-the-art results. Additionally, we developed a web server based on the best-performing models to predict NOVA food processing labels using nutrient concentration of a food product.

For nutrient panel-based models, the best performances were achieved with LGBM using 102 nutrients, Random Forest using 65 nutrients, and Gradient Boost using 13 nutrients. Further for NLP-based models, where we combined textual columns ('Food Description', 'Category Name', and 'Macro Class') and used their embeddings, the highest accuracy was achieved with XLM-Roberta combined with Gradient Boost using 102 nutrients, LGBM combined with DistilBERT using 65 nutrients, and LGBM combined with GPT-2 using 13 nutrients. Furthermore, we utilized SHAP analysis to identify key nutrients contributing to the model's predictive capability, as well as features that are essential in determining the processing level of food products.

In summary, this work represents significant advancement in the prediction of food processing, highlighting the potential of combining nutrient panels with advanced machine learning and NLP techniques. Future research should focus on expanding and diversifying datasets and further refining NLP methods to enhance model performance and generalizability.

## 6. Funding Source

The current work has not received any specific grant from any funding agencies.

## 7. Conflict of interest

## 8. Authors' contributions

GB conceived the idea, designed the methodology, and supervised the project. NA collected the data, conducted the experiments, developed the models, and performed the literature survey. SB and RD implemented the webserver. NA and GB drafted the manuscript, proofread and approved the final manuscript.

## 9. Acknowledgments

GB thanks Indraprastha Institute of Information Technology Delhi (IIIT-Delhi) for the computational support. GB thanks Technology Innovation Hub (TiH) Anubhuti for the research grant. NA is an undergraduate research intern in the Complex Systems Laboratory and is thankful for the lab resources. This study was supported by the Infosys Center for Artificial Intelligence and Centre of Excellence in Healthcare at IIIT-Delhi.

## 10. Data Availability Statement

The complete dataset of food ingredients is available on [GitHub](GitHub).

**SUPPLEMENTARY DATA 1**

# 1. Comparison of various food processing classification systems

| System | Categories' Description |
|---|---|
| NOVA | 1. Unprocessed and minimally processed: Unprocessed foods, sourced directly from nature, include fruits, vegetables, meat, eggs, milk, fungi, algae, and water. Minimally processed foods, such as dried fruits, ground spices, pasteurized milk, and vacuum-packaged vegetables, undergo basic alterations while retaining their natural state.<br>2. Processed culinary ingredients: Processed culinary ingredients like oils, butter, sugar, and salt are derived from natural sources through methods such as pressing and refining.<br>3. Processed foods: These are produced by adding ingredients such as oil, sugar, or salt to whole foods, enhancing their shelf life and improving their taste and visual appeal. Examples of such processed foods include canned and bottled vegetables in brine, salted nuts, and cheese.<br>4. Ultra-processed foods: Ultra-processed foods are mainly formulated from or entirely composed of ingredients sourced from foods. These products undergo diverse techniques, including hydrogenation, hydrolysis, extrusion, molding, reshaping, and preliminary treatments such as frying or baking. Examples include burgers, hot dogs, cookies, and energy bars. |
| IFIC-JTF | 1. Minimally processed: Foods that undergo minimal processing or production, preserving most of their natural properties, include milk, coffee, fruits, vegetables, meat, and eggs.<br>2. Foods processed for preservation: Processed foods designed to maintain and enhance food's nutritional content and freshness at its optimal state include fruit juices and cooked, canned, or frozen vegetables and fruits.<br>3. Mixtures of combined ingredients: Food items incorporating sweeteners, spices, oils, coloring agents, flavorings, and preservatives to enhance safety, flavor, and |
| | visual appeal. Examples include breads, sugars, cheeses, condiments, and tacos or tortillas.<br>4. Ready-to-eat processed: Foods requiring minimal or no preparation, categorized into 'packaged ready-to-eat foods' and 'mixtures possibly requiring preparation.' Examples include soft drinks, candies, savory snacks, cereals, deli meats, and alcoholic beverages.<br>5. Prepared foods/meals: Foods packaged for convenience and freshness, facilitating easy preparation. Examples include pizza, pre-cooked meat dishes, pasta, and pre-made meals. |
| UNC | 1. Less processed<br>    a. Unprocessed & minimally processed: Foods consisting of a single ingredient with minimal or negligible alterations that preserve their natural properties. Examples include plain milk, fresh, frozen, dried plain fruits or vegetables, unseasoned eggs and meat, whole grain flour, honey, herbs, and spices.<br>2. Basic processed<br>    a. Processed essential ingredients: Individual food components isolated through extraction or purification via physical or chemical processes, altering the inherent properties of the food. Examples include whole-grain pasta oil, unsalted butter, sugar, and salt.<br>    b. Processed for basic preservation or precooking: Single minimally processed foods subject to physical or chemical modifications aimed at preservation or pre-cooking while maintaining their integrity as individual food items. Examples include unsweetened/unflavored canned fruit, vegetables, and legumes; plain peanut butter; refined grain pasta, white rice; and plain yogurt.<br>3. Moderately processed<br>    a. Moderately processed for flavor: Single minimally or moderately processed foods augmented with flavor additives to enhance taste. |
|  | Examples include sweetened fruit juice, flavored milk, frozen French fries, salted peanut butter, smoked or cured meats, and cheese.<br>b. Moderately processed grain products: Grain products crafted from whole-grain flour combined with water, salt, and/or yeast. Examples include whole grain breads, tortillas, or crackers with no added sugar or fat.<br>4. Highly processed<br>a. Highly processed ingredients: Multi-ingredient industrially formulated mixtures are processed to the extent that they lose their original plant or animal identity. Examples include tomato sauce, salsa, mayonnaise, salad dressing, and ketchup.<br>b. Highly processed stand-alone: Multi-ingredient industrially formulated mixtures are processed to the extent that they lose their original plant or animal identity. Examples include soda, fruit drinks, bread made with refined flour, pastries, ice cream, and candy. |
| IARC-EPIC | 1. Non-processed: Foods are eaten raw without additional processing or preparation except for washing, cutting, or squeezing. Examples include Raw fruits, non-processed nuts, fresh raw vegetables, fresh grated vegetables, fresh juices, fresh and not enriched farmer's milk, raw meat, raw egg whites, and honey.<br>2. Modestly or moderately processed:<br>a. Industrial and commercial foods that undergo limited processing are consumed without additional cooking. Examples include dried or semi-dried fruits, nuts and seeds, frozen or vacuum-packed raw meat, extra virgin olive oil, and fruits.<br>b. Foods processed at home and cooked or prepared from raw or minimally processed ingredients. Examples include fresh vacuum-packed or frozen cooked potato (including homemade French fries); cooked fruit; fresh or frozen cooked vegetables, dried boiled legumes, boiled grain; whole-meal boiled rice, fresh or vacuum-packed cooked meat; and fish.<br>3. Processed: Industrially prepared foods undergo extensive processing methods like drying, flaking, and deep frying, often containing industrial ingredients. Examples |

| | | |
|---|---|---|
| | | include bakery and catering foods, requiring minimal domestic preparation such as heating or cooking. This category is divided into staple/basic and highly processed foods. Examples include:<br>a. Processed staple/basic: Bread, rice, milk, butter, vegetable oils.<br>b. Highly processed: Cakes, biscuits, breakfast cereals, crisp bread, processed meat, fish, yogurt, and cheese. |
| | NIPH | 1. Modern industrialized: Foods integrated into the Mexican diet, either as standalone items or combined with other ingredients, making separation impossible. Examples include Powdered milk, non-fat milk, 1% milk, breakfast cereals, whole wheat bread, salty wheat bread, sausages, granulated and liquid sugar, sweetened drinks, instant coffee, baby formulas, compotes, and supplements.<br>2. Industrialized traditional: Foods deeply rooted in conventional Mexican culinary culture, dating back to customs and traditions before the 20th century, now mass-produced industrially. Examples include Corn flour for tortillas or atoles and whole cow milk.<br>3. Non-industrialized:<br>a. Contemporary preparations beyond home settings featuring ingredients not customary in Mexican cuisine. Examples include modern preparations outside the home, such as burgers, sandwiches, pizza, and milkshakes.<br>b. Traditional preparations beyond home settings characterized by ingredients difficult to distinguish. Typically prepared locally or domestically and entrenched in Mexico's culinary heritage. Examples include Beans or stews with beans, tacos, atole, tamales, fresh water, broths, salsas, fish, meat stews, fried fish, soups, and salads.<br>c. Locally crafted traditional foods embodying authentic Mexican cuisine, typically homemade or artisanal on a small scale. Examples include corn |

| | |
|---|---|
| | tortillas, salty and sweet bread (bolillo), animal fats such as pig skin or lard, and homemade sugar and drinks.<br>d. Unprocessed raw foods are subjected only to collection, selection, and cleaning without further processing. Examples include fruits, vegetables, legumes, cereals, tubers, red and white meats, fish, and eggs. |
| IFPRI | 1. Unprocessed: undefined. Examples include staple foods like corn and other grains, roots and tubers, beans, vegetables, fruits, meat, fish, eggs, and dairy, including fresh, dried milk cream.<br>2. Primary or partially processed: undefined. Examples include corn and dairy products like evaporated milk, cheese, yogurt, and animal fats.<br>3. Highly processed: Foods subjected to secondary processing to make them readily edible, often containing elevated levels of added sugars, fats, or salt. Examples include pastries, cookies, crackers, sausage, prepared meats, ice cream, frozen desserts, breakfast cereals, confectionery (sweets, chocolate), fat spreads and shortening, pasta products, soft drinks, formula, and complementary foods. |

**Table S1: Comprehensive comparison of existing processing classification systems and corresponding categories**

## 2. Comparison of food products across NOVA classes

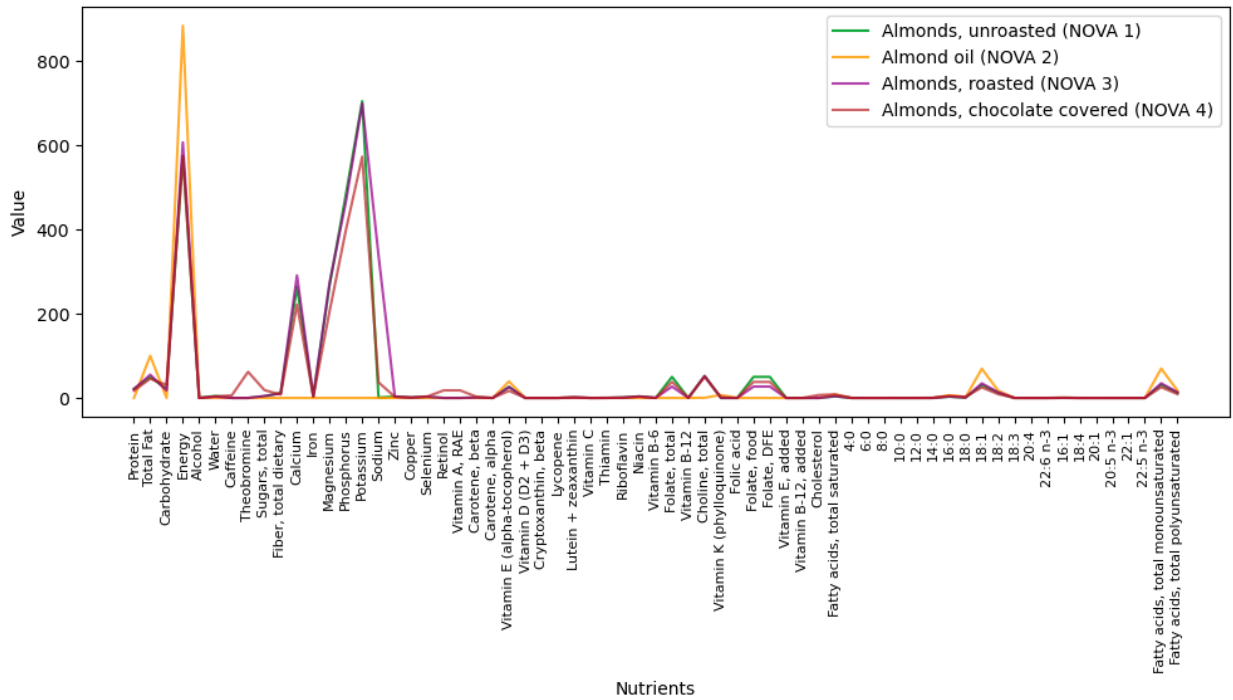

**Figure S1: Nutrient concentration comparison across NOVA classes of Almond as the core ingredient.** There is a difference in the nutrient concentration of food products with almonds as the core ingredient across the 4 NOVA classes.

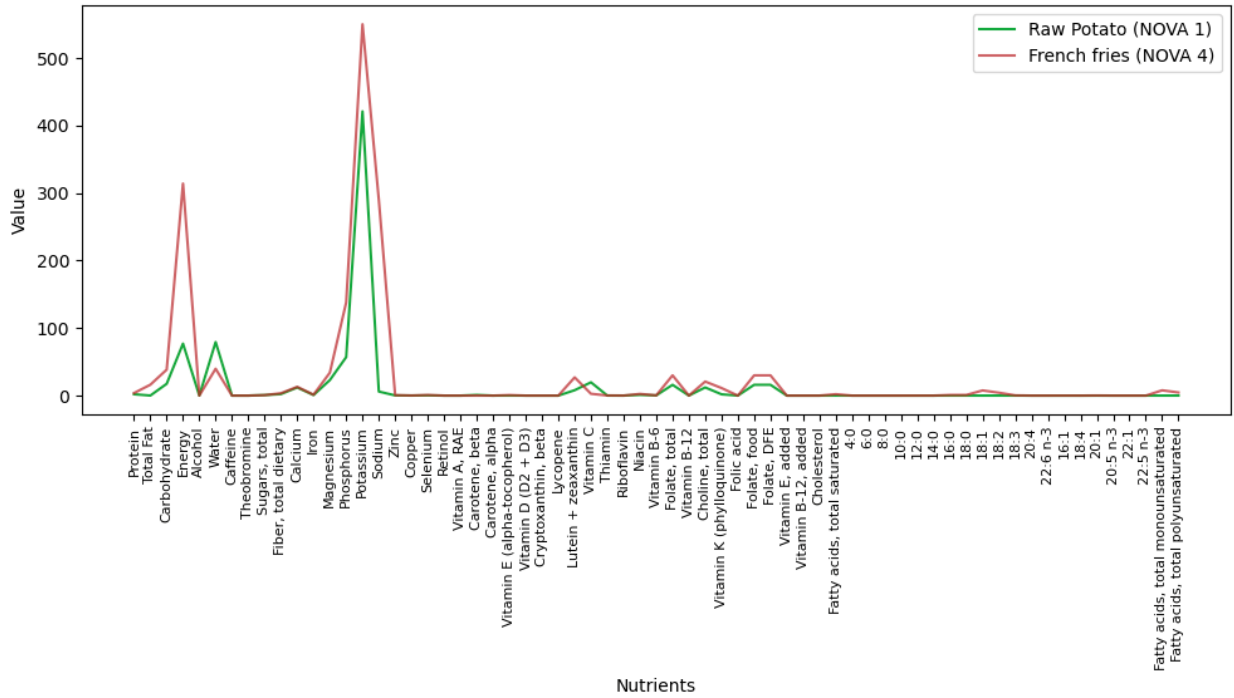

**Figure S2: Nutrient concentration comparison across NOVA Class 1 and NOVA Class 4 of Potato as the core ingredient.** There is a difference in the nutrient concentration of food products with potatoes as the core ingredient across NOVA Class 1 and NOVA Class 4. The concentration of nutrients in french fries follows a trend, with it being higher than in raw potatoes.

## 3. **Nutrients utilized in this study**

| | | |
|---|---|---|
| Protein | Vitamin B-12 | Cyanidin |
| Total Fat | Choline, total | Petunidin |
| Carbohydrate | Vitamin K (phylloquinone) | Delphinidin |
| Energy | Folic acid | Malvidin |
| Alcohol | Folate, food | Pelargonidin |
| Water | Folate, DFE | Peonidin |
| Caffeine | Vitamin E, added | (+)-Catechin |
| Theobromine | Vitamin B-12, added | (-)-Epigallocatechin |
| Sugars, total | Cholesterol | (-)-Epicatechin |
| Fiber, total dietary | Fatty acids, total saturated | (-)-Epicatechin 3-gallate |
| Calcium | 4:0 | (-)-Epigallocatechin 3-gallate |
| Iron | 6:0 | Theaflavin |
| Magnesium | 8:0 | Thearubigins |
| Phosphorus | 10:0 | Eriodictyol |
| Potassium | 12:0 | Hesperetin |
| Sodium | 14:0 | Naringenin |
| Zinc | 16:0 | Apigenin |
| Copper | 18:0 | Luteolin |
| Selenium | 18:1 | Isorhamnetin |
| Retinol | 18:2 | Kaempferol |
| Vitamin A, RAE | 18:3 | Myricetin |
| Carotene, beta | 20:4 | Quercetin |
| Carotene, alpha | 22:6 n-3 | Theaflavin-3,3'-digallate |
| Vitamin E (alpha-tocopherol) | 16:1 | Theaflavin-3'-gallate |
| Vitamin D (D2 + D3) | 18:4 | Theaflavin-3-gallate |
| Cryptoxanthin, beta | 20:1 | (+)-Gallocatechin |
| Lycopene | 20:5 n-3 | Total flavonoids |
| Lutein + zeaxanthin | 22:1 | Total anthocyanidins |
| Vitamin C | 22:5 n-3 | Total catechins (monomeric flavan-3- |

|  |  | ols only) |
|---|---|---|
| Thiamin | Fatty acids, total monounsaturated | Total flavan-3-ols |
| Riboflavin | Fatty acids, total polyunsaturated | Total flavanones |
| Niacin | Daidzein | Total flavones |
| Vitamin B-6 | Genistein | Total flavonols |
| Folate, total | Glycitein | Total isoflavones |

**Table S2: Dataset of 102 nutrients utilized in the study**

| Protein | Folate, total |
|---|---|
| Total Fat | Vitamin B-12 |
| Carbohydrate | Choline, total |
| Energy | Vitamin K (phylloquinone) |
| Alcohol | Folic acid |
| Water | Folate, food |
| Caffeine | Folate, DFE |
| Theobromine | Vitamin E, added |
| Sugars, total | Vitamin B-12, added |
| Fiber, total dietary | Cholesterol |
| Calcium | Fatty acids, total saturated |
| Iron | 4:0 |
| Magnesium | 6:0 |
| Phosphorus | 8:0 |
| Potassium | 10:0 |
| Sodium | 12:0 |
| Zinc | 14:0 |
| Copper | 16:0 |
| Selenium | 18:0 |
| Retinol | 18:1 |
| Vitamin A, RAE | 18:2 |
| Carotene, beta | 18:3 |
| Carotene, alpha | 20:4 |
| Vitamin E (alpha-tocopherol) | 22:6 n-3 |
| Vitamin D (D2 + D3) | 16:1 |
| Cryptoxanthin, beta | 18:4 |
| Lycopene | 20:1 |
| Lutein + zeaxanthin | 20:5 n-3 |
| Vitamin C | 22:1 |
| Thiamin | 22:5 n-3 |
| Riboflavin | Fatty acids, total monounsaturated |
| Niacin | Fatty acids, total polyunsaturated |
| Vitamin B-6 | |

**Table S3: Dataset of 65 nutrients utilized in the study**

| Nutrients |
|---|
| Protein |
| Total Fat |
| Carbohydrate |
| Sugars, total |
| Fiber, total dietary |
| Calcium |
| Iron |
| Sodium |
| Vitamin D (D2 + D3) |
| Cholesterol |
| Fatty acids, total saturated |
| Potassium |
| Energy |

**Table S4: Dataset of 13 nutrients utilized in the study.** The 13-nutrient panel consists of all the nutrients mandated by the FDA except for Trans Fat since it was not available in the FNDDS data.

4. Exploratory Data Analysis

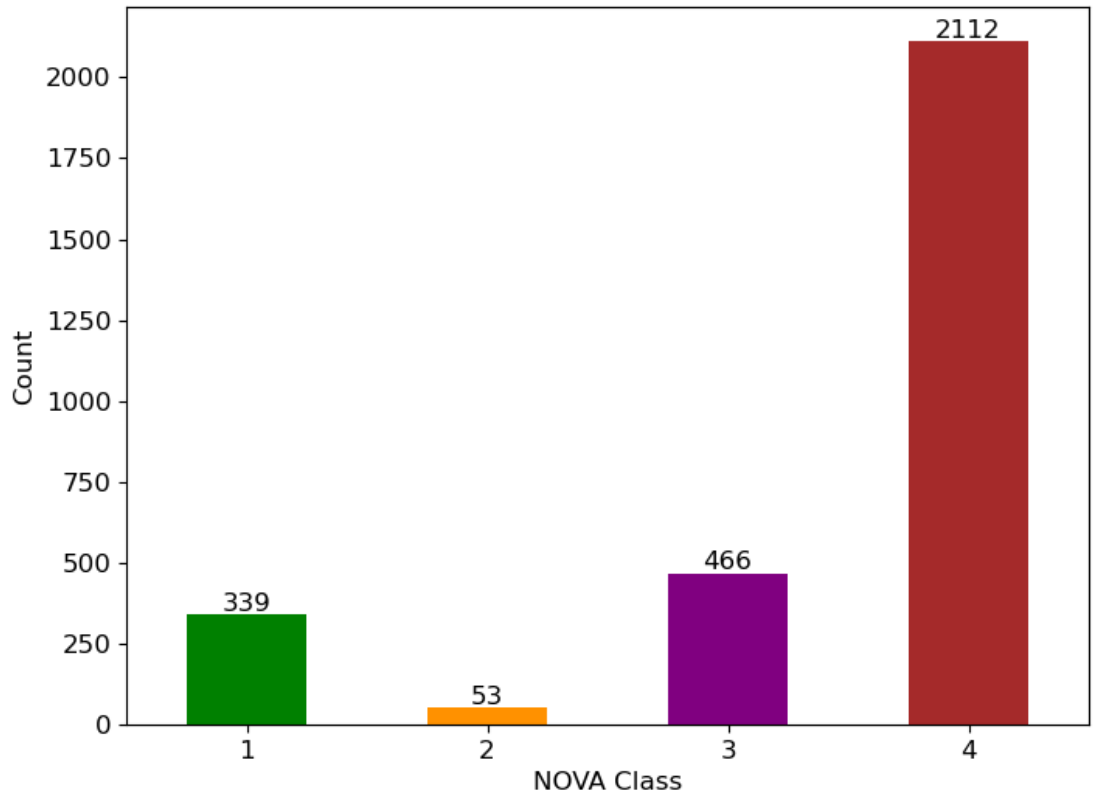

**Figure S3: Distribution of 2970 FNDDS food products across NOVA classes.**

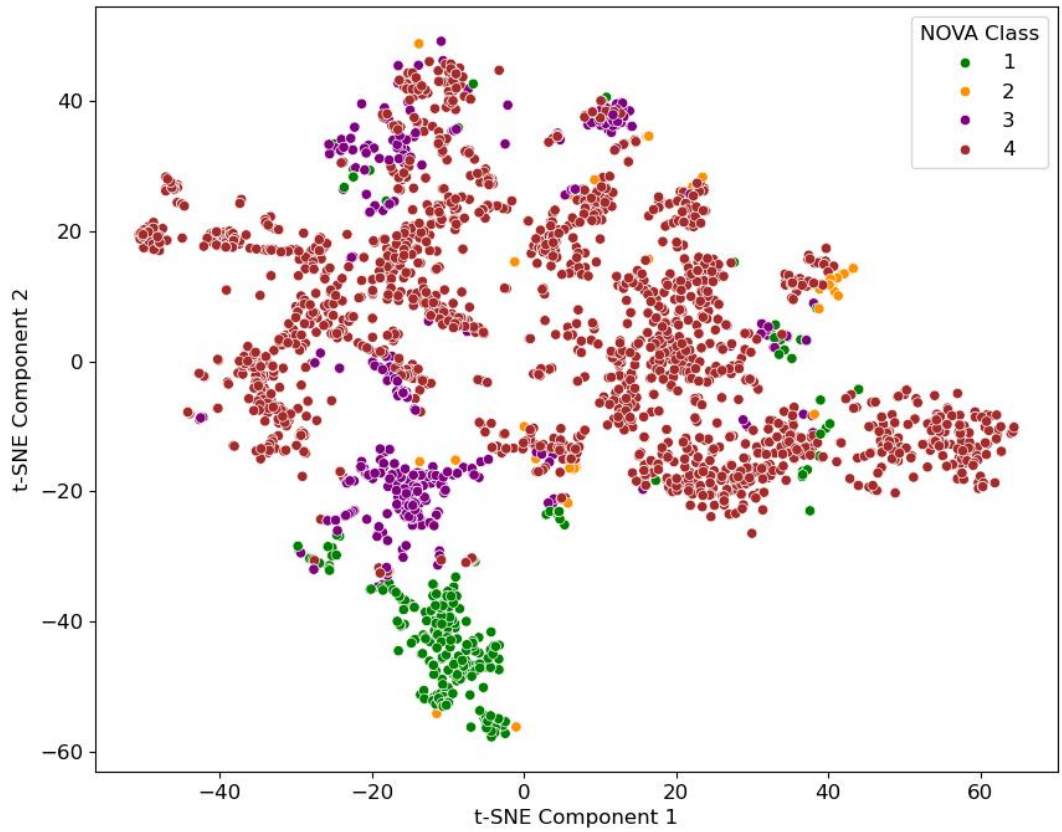

**Figure S4: The t-SNE plot depicting clusters of FNDDS food products across NOVA classes.**

## 5. ROC and Precision Recall Curves for 102 nutrients

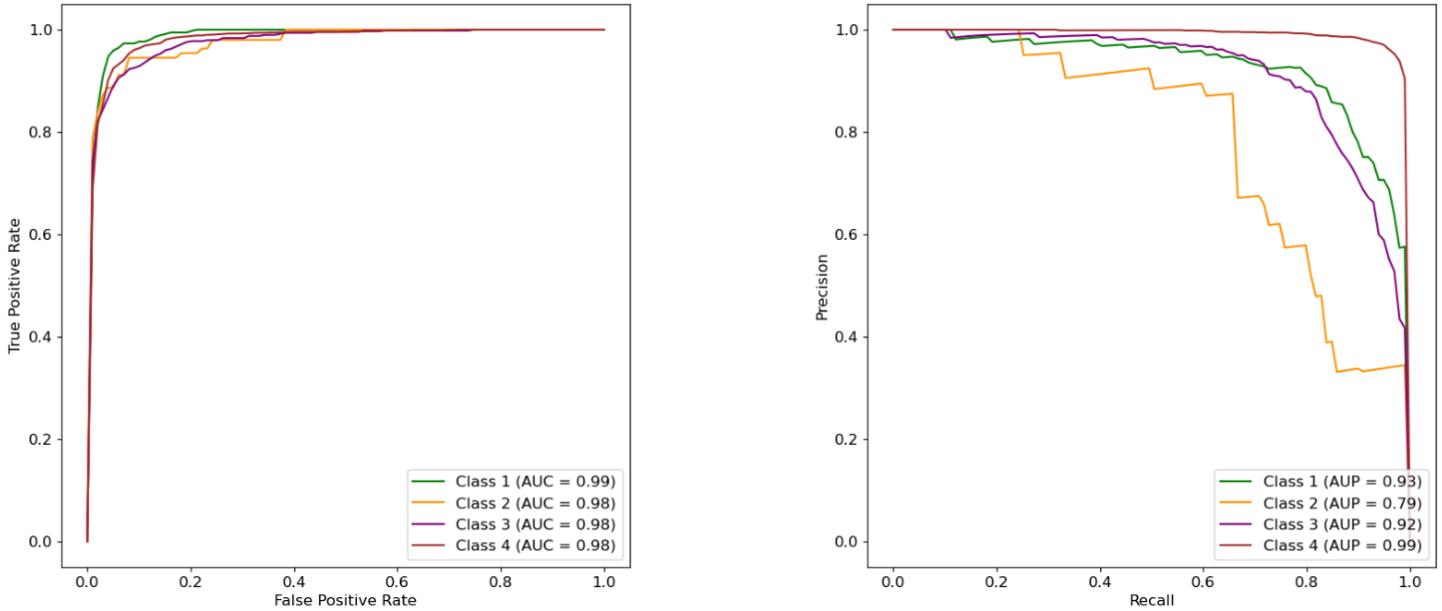

**Figure S5: ROC and Precision-Recall curves for 102 nutrients.** LGBM with SMOTE attained the highest AUC scores of 0.99, 0.98, 0.98, and 0.98 for NOVA Classes 1, 2, 3, and 4, respectively. Additionally, it achieved AUP scores of 0.93, 0.79, 0.92, and 0.99 for NOVA Classes 1, 2, 3, and 4, respectively.

## 6. ROC and Precision Recall Curves for 65 nutrients

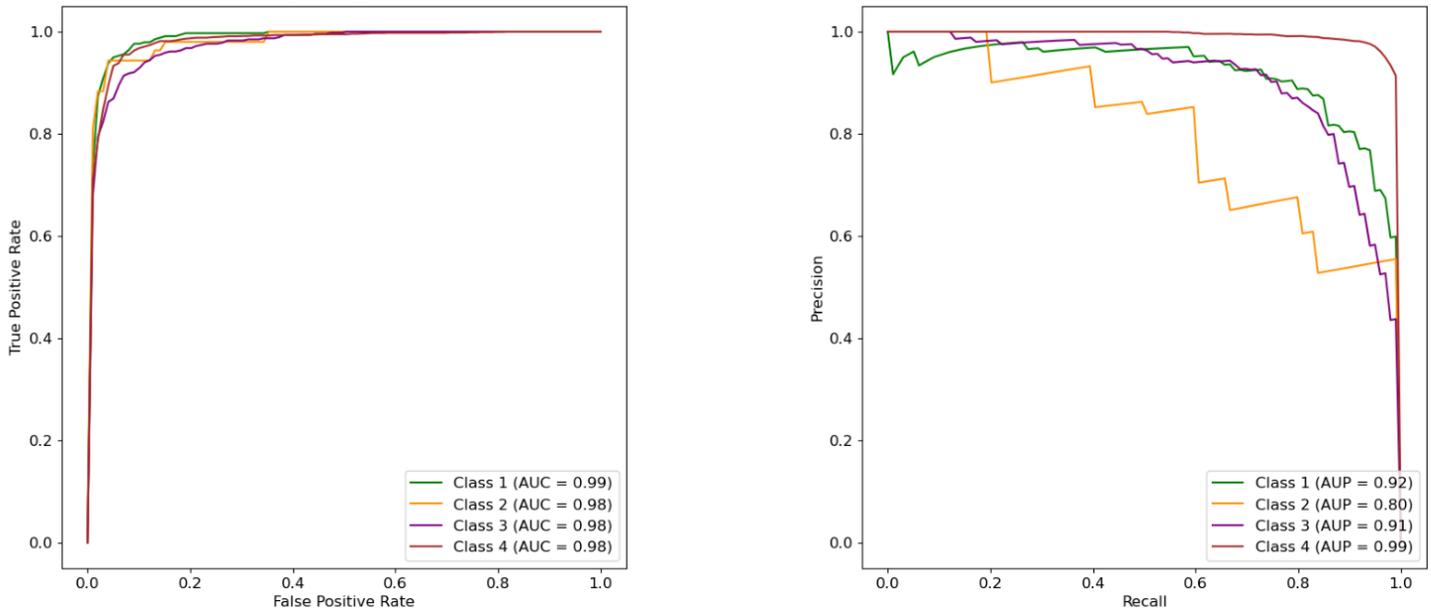

**Figure S6: ROC and Precision-Recall curves for 65 nutrients.** LGBM with stratified k-fold, attained the highest AUC scores of 0.99, 0.98, 0.98, and 0.98 for NOVA Classes 1, 2, 3, and 4, respectively. Additionally, it achieved AUP scores of 0.92, 0.80, 0.91, and 0.99 for NOVA Classes 1, 2, 3, and 4, respectively.

## 7. ROC and Precision Recall Curves for 13 nutrients

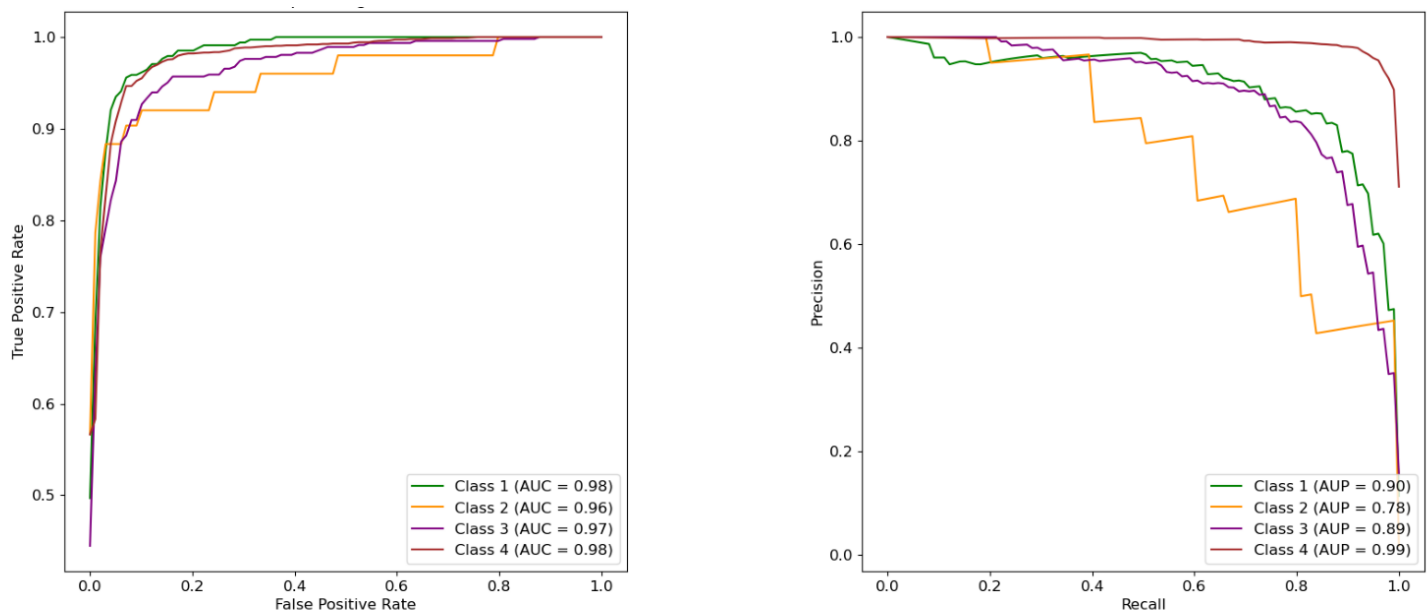

**Figure S7: ROC and Precision-Recall curves for 13 nutrients.** Gradient Boost with SMOTE and stratified k-fold, attained the highest AUC scores of 0.98, 0.96, 0.97, and 0.98 for NOVA Classes 1, 2, 3, and 4, respectively. Additionally, it achieved AUP scores of 0.90, 0.78, 0.89, and 0.99 for NOVA Classes 1, 2, 3, and 4, respectively.

## 8. Comparison of model performance with previous studies

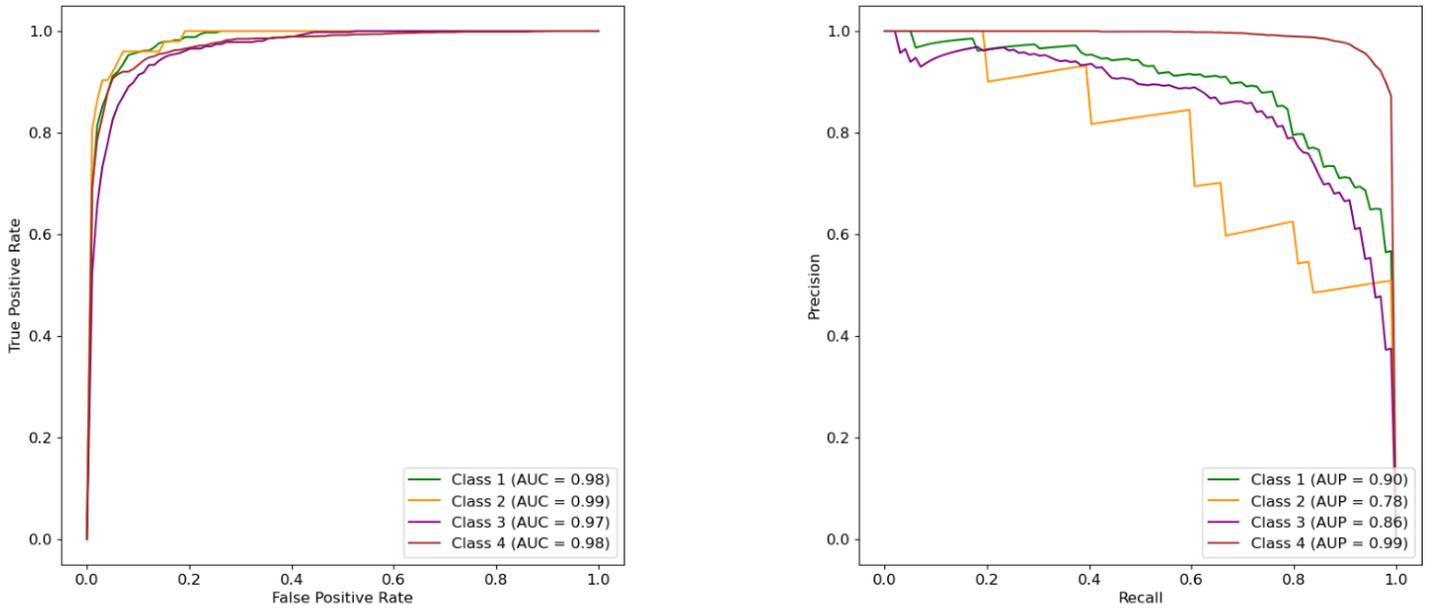

**Figure S8: ROC and Precision-Recall curves for 12 nutrients.** The Random Forest Classifier attained AUC scores of 0.98, 0.99, 0.97, and 0.98 for NOVA Classes 1, 2, 3, and 4, respectively. Additionally, it achieved AUP scores of 0.90, 0.78, 0.86, and 0.99 for NOVA Classes 1, 2, 3, and 4, respectively.

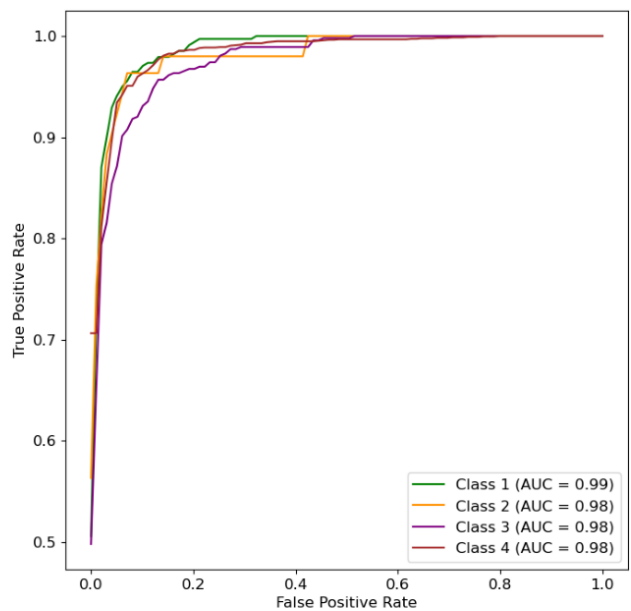 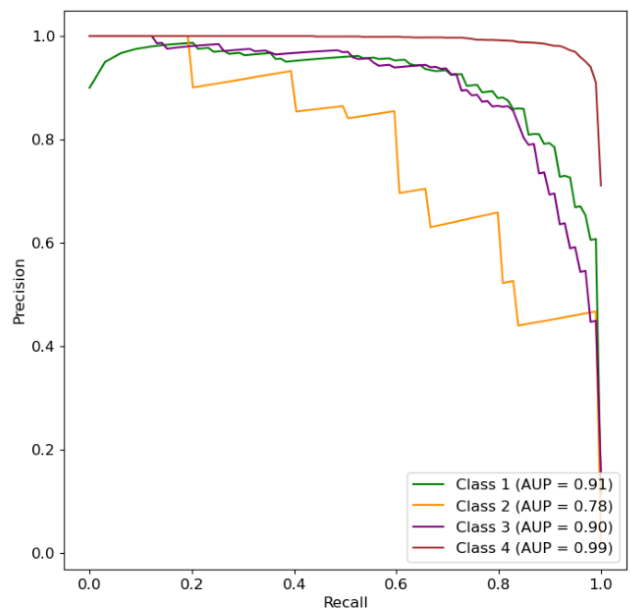

**Figure S9: ROC and Precision-Recall curves for 62 nutrients.** The LGBM Classifier attained AUC scores of 0.99, 0.98, 0.98, and 0.98 for NOVA Classes 1, 2, 3, and 4, respectively. Additionally, it achieved AUP scores of 0.91, 0.78, 0.90, and 0.99 for NOVA Classes 1, 2, 3, and 4, respectively.

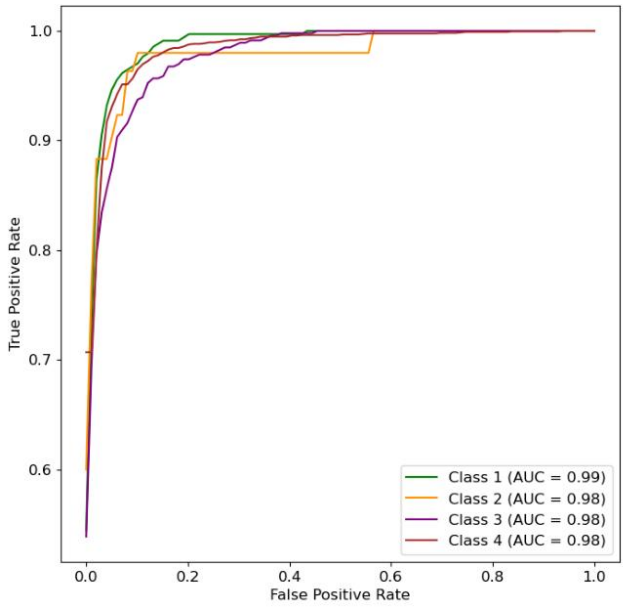 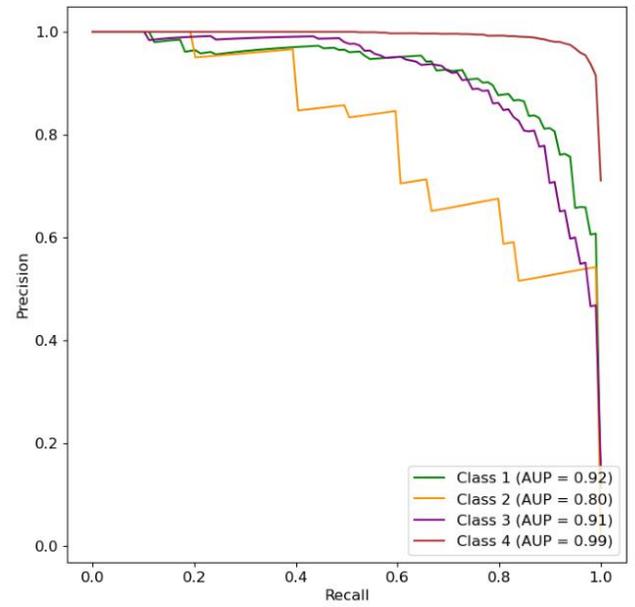

**Figure S10: ROC and Precision-Recall curves for 99 nutrients.** The LGBM Classifier attained AUC scores of 0.99, 0.98, 0.98, and 0.98 for NOVA Classes 1, 2, 3, and 4, respectively. Additionally, it achieved AUP scores of 0.92, 0.80, 0.91, and 0.99 for NOVA Classes 1, 2, 3, and 4, respectively.

# 9. Feature Importance

## 9.1 Feature importance of 65 nutrients

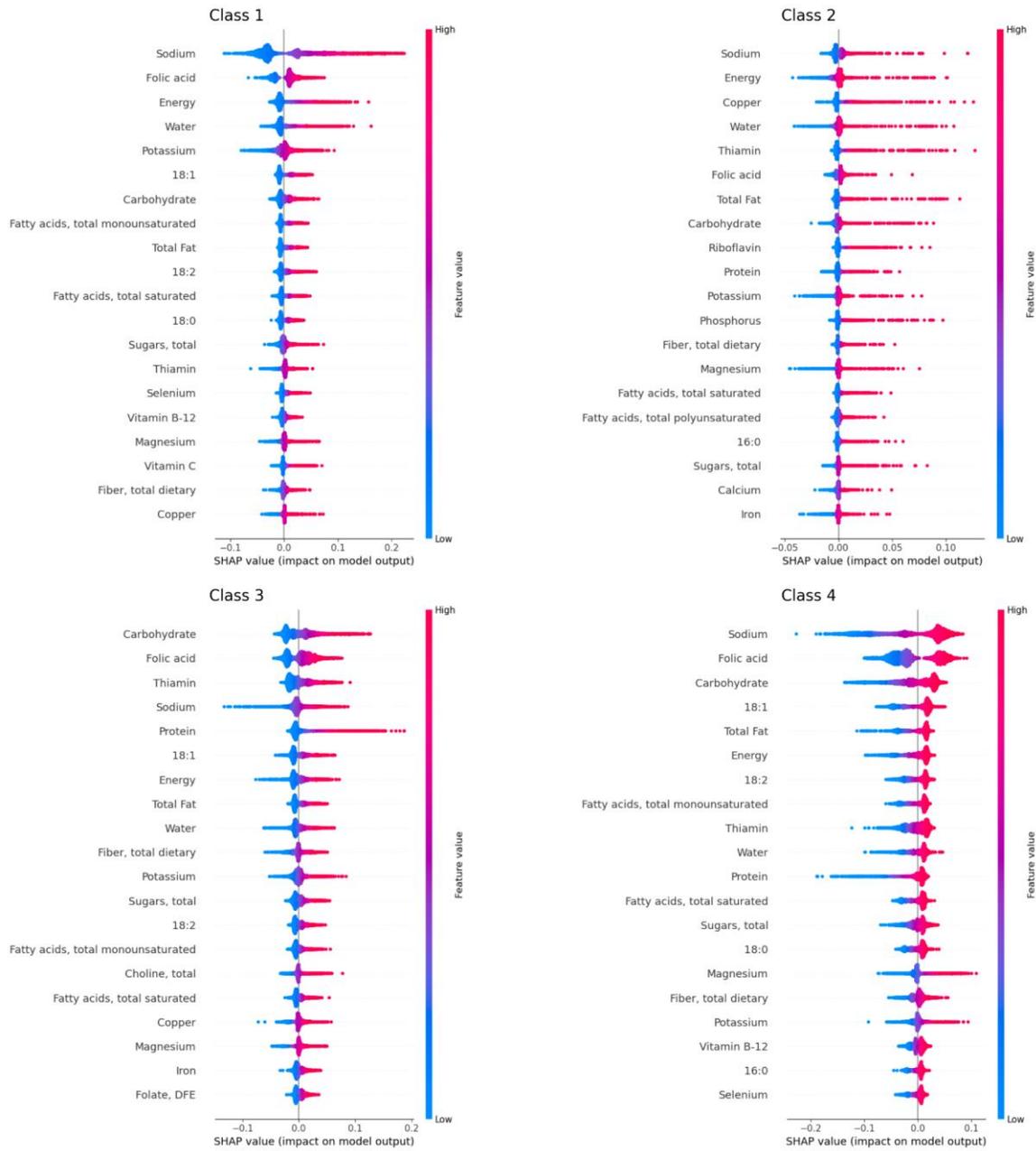

**Figure S11: SHAP plots for 65 nutrient panel.** In NOVA Class 1, Sodium, Folic Acid, Energy, Water, and Potassium are the top 5 features. In Class 2, Sodium, Energy, Copper, Water, and Thiamin are the top 5 features. The top 5 most essential features for Class 3 are Carbohydrates,

Folic Acid, Thiamin, Sodium, and Protein. In Class 4 Sodium, Folic Acid, Carbohydrate, 18:1(oleic acid), and Total Fat emerge as the top 5 features.

## 9.2 Feature importance of 13 nutrients

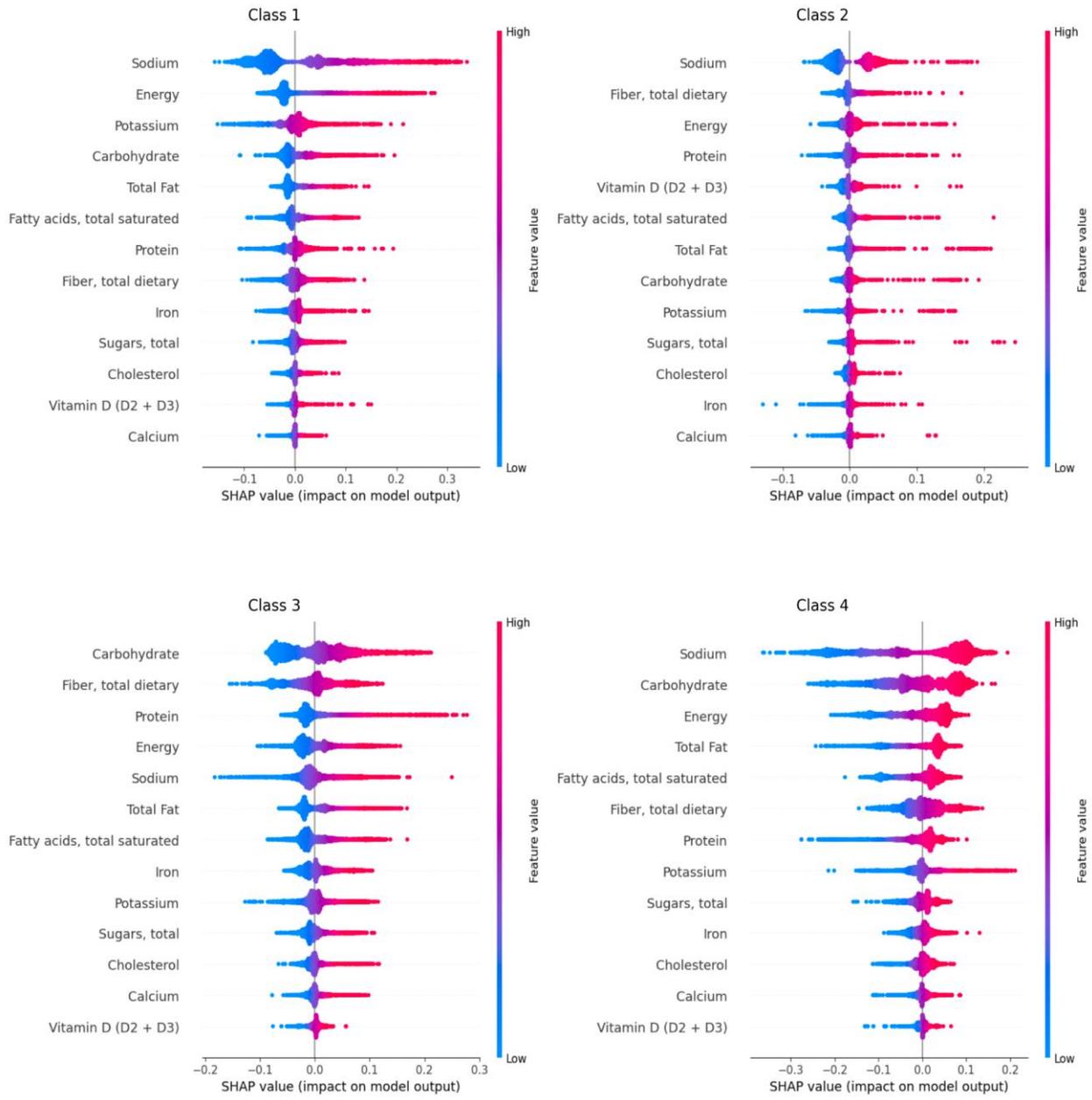

**Figure S12: SHAP plots for 13 nutrient panel.** In NOVA Class 1, Sodium, Energy, and Potassium are the top 3 features. In Class 2, Sodium, Fiber, and Energy are the top 3 features. The top 3 most essential features for Class 3 are Carbohydrates, Fiber, and Protein. In Class 4 Sodium, Carbohydrate, and Energy emerge as the top 3 features

# 10. Analysis incorporating the merger and exclusion of NOVA Class 2 and wide-ranging evaluation metrics

## 10.1 Methods and materials.

In one of the experiments carried out as part of our study, we opted to consolidate NOVA Class 2 with Class 3 due to the limited number of food items in the former, totaling only 53 items. This consolidation aimed to assess the performance of the model when integrating Class 2 with Class 3. Additionally, we conducted another experiment wherein we removed NOVA Class 2 to evaluate its impact on our results. This investigation allowed us to gauge the influence of Class 2 on the overall outcome.

## 10.2 Results

### 10.2.1 Merger of NOVA Class 2

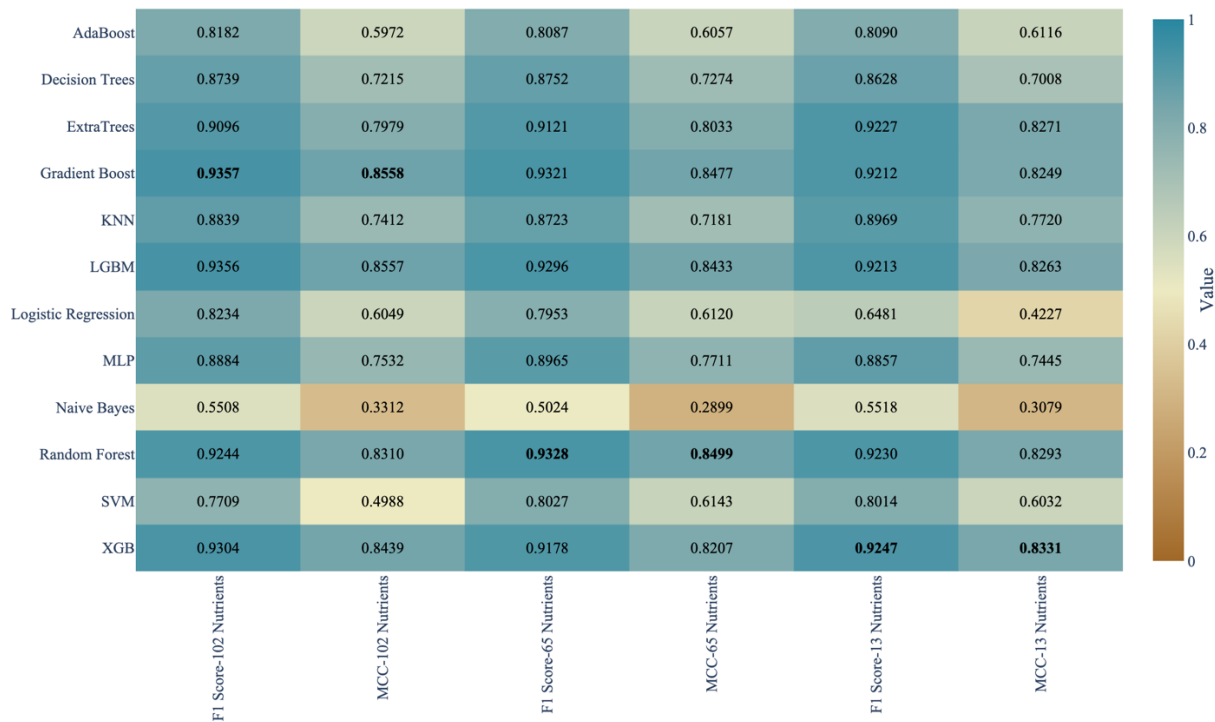

**Figure S13: Model performance for the integration of NOVA Class 2 and NOVA Class 3 across nutrient panels. The best scores for each nutrient panel are highlighted in bold.**

Given that NOVA Class 2 comprises only 1.8% of the whole data, we evaluated the impact of either avoiding it or merging it with NOVA Class 3. This experiment tells us about the impact of Class 2 from the classification point of view. The performance metrics of various machine learning models trained on a dataset where 'NOVA Class 2 was integrated with NOVA Class 3' are presented in Figure S16. Firstly, for the 102-nutrient panel, Gradient Boost and LGBM returned the highest predictive capability, with F1-scores of 0.9357 and 0.9356, respectively, along with MCC scores of 0.8558 and 0.8557, respectively. Refer to Figure S14 for the best ROC and precision-recall curves for 102 nutrients. Secondly, for the 65-nutrient panel, Random Forest presented the best performance with an F1-score of 0.9328 and an MCC of 0.8499. Refer to Figure

S15 for the best ROC and precision-recall curves for 65 nutrients. Finally, for the 13 nutrients panel, XGB yielded the highest predictive capability with an F1-score of 0.9247 and an MCC of 0.8331. Refer to Figure S16 for the best ROC and precision-recall curves for 13 nutrients.

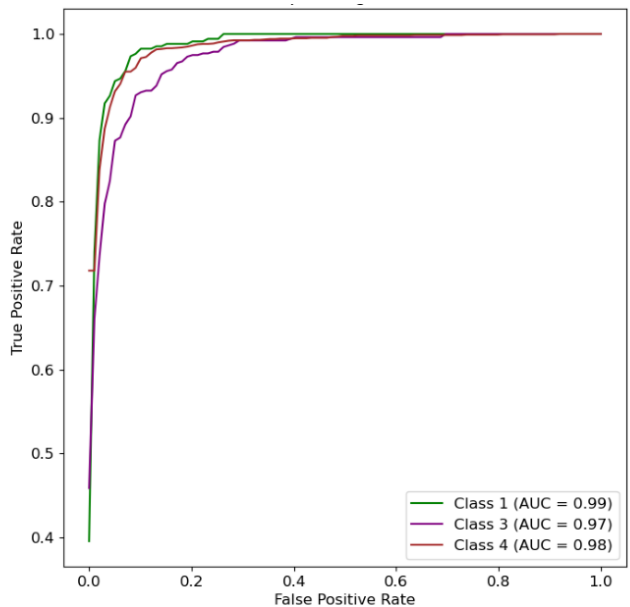 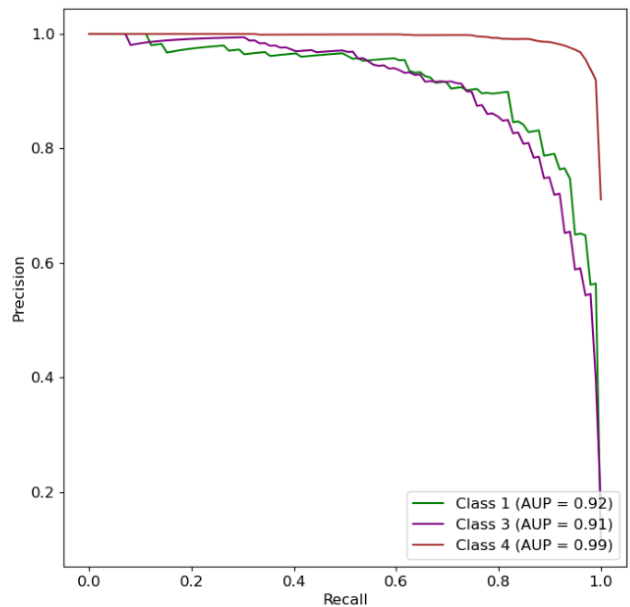

**Figure S14: ROC and Precision-Recall curves for 102 nutrients.** The Gradient Boost Classifier attained highest AUC scores of 0.99, 0.97, and 0.98 for NOVA Classes 1, 3, and 4, respectively. Additionally, it achieved AUP scores of 0.92, 0.91, and 0.99 for NOVA Classes 1, 3, and 4, respectively.

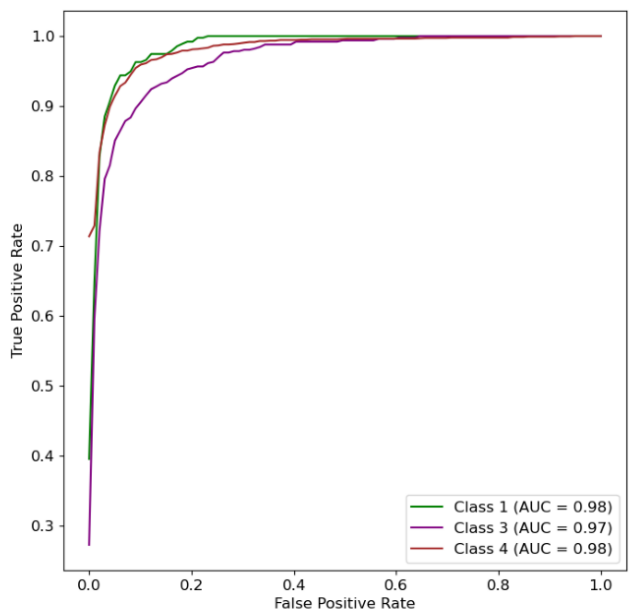 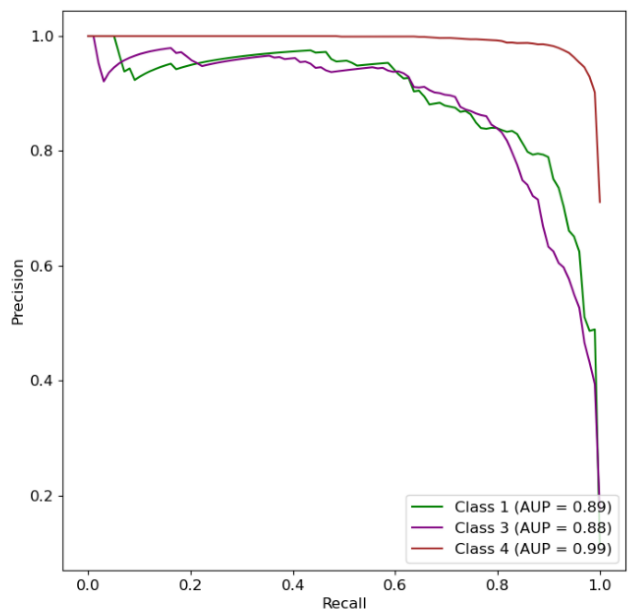

**Figure S15: ROC and Precision-Recall curves for 65 nutrients.** The Random Forest Classifier attained highest AUC scores of 0.98, 0.97, and 0.98 for NOVA Classes 1, 3, and 4, respectively. Additionally, it achieved AUP scores of 0.89, 0.88, and 0.99 for NOVA Classes 1, 3, and 4, respectively.

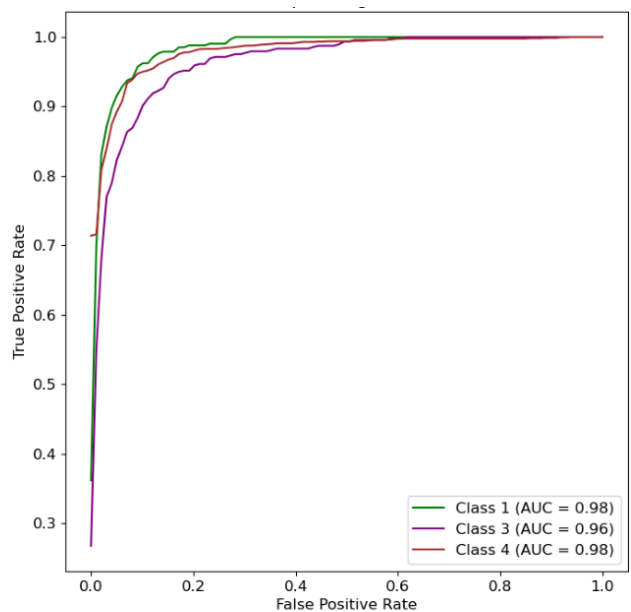 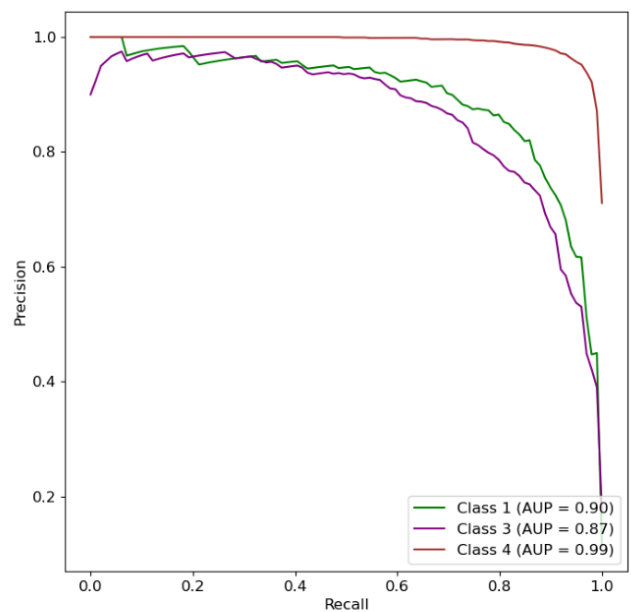

**Figure S16: ROC and Precision-Recall curves for 13 nutrients.** The XGB classifier attained the highest AUC scores of 0.98, 0.96, and 0.98 for NOVA Classes 1, 3, and 4, respectively. Additionally, it achieved AUP scores of 0.90, 0.87, and 0.99 for NOVA Classes 1, 3, and 4, respectively.

## 10.2.2 Removal of NOVA Class 2

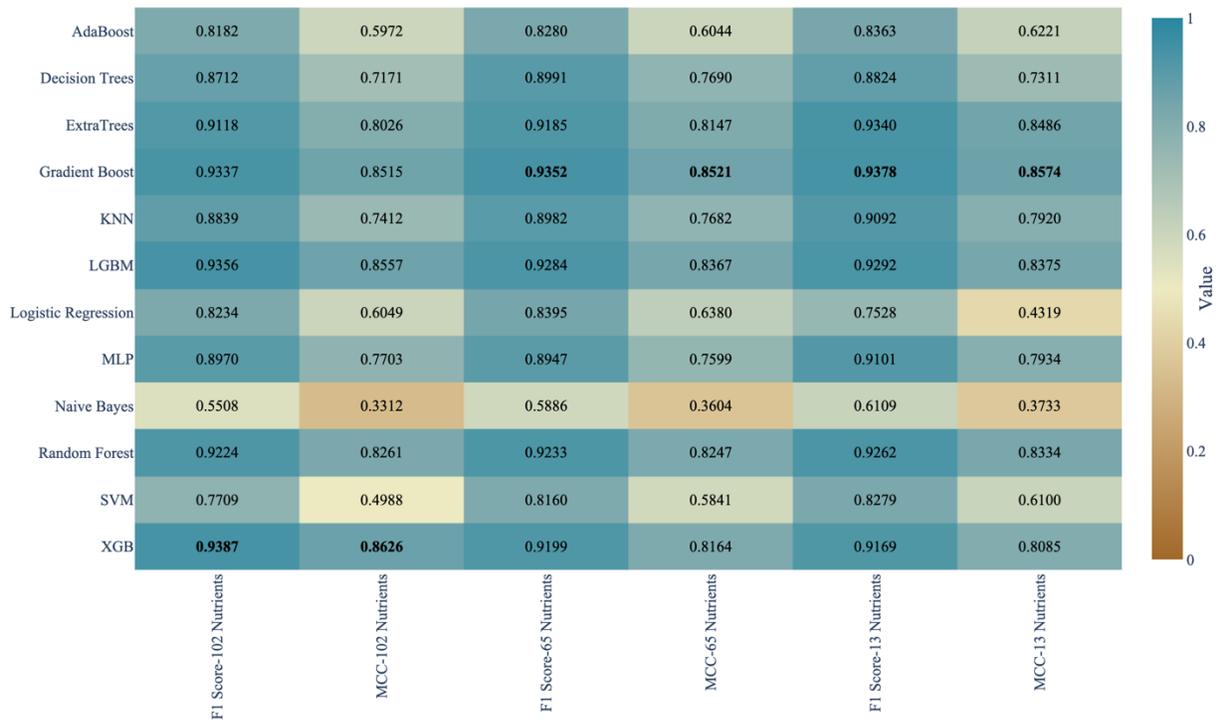

**Figure S17: Model performance for the removal of NOVA Class 2 across nutrient panels. The best scores for each nutrient panel are highlighted in bold.**

The performance metrics of various machine learning models trained on a dataset where 'NOVA Class 2 was removed' are presented in Figure S20. Firstly, for the 102-nutrient panel, XGB yielded the highest predictive capability, achieving F1-scores of 0.9387, along with MCC scores of 0.8626. Refer to Figure S18 for the best ROC and precision-recall curves for 102 nutrients. Secondly, in the case of the 65-nutrient panel, Gradient Boost exhibits the highest predictive capability with an F1-score of 0.9352 and an MCC of 0.8521. Refer to Figure S19 for the best ROC and precision-recall curves for 65 nutrients. Finally, for the 13 nutrients panel, Gradient Boost also achieves the highest predictive capability with an F1-score of 0.9378 and an MCC of 0.8574. Refer to Figure S20 for the best ROC and precision-recall curves for 13 nutrients.

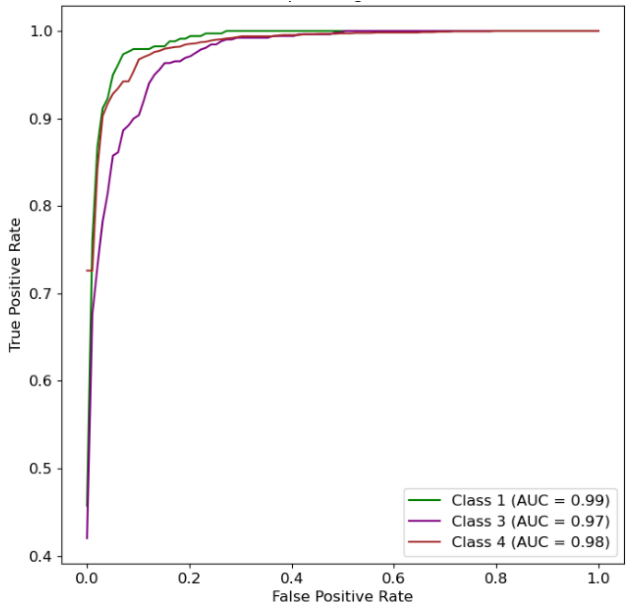 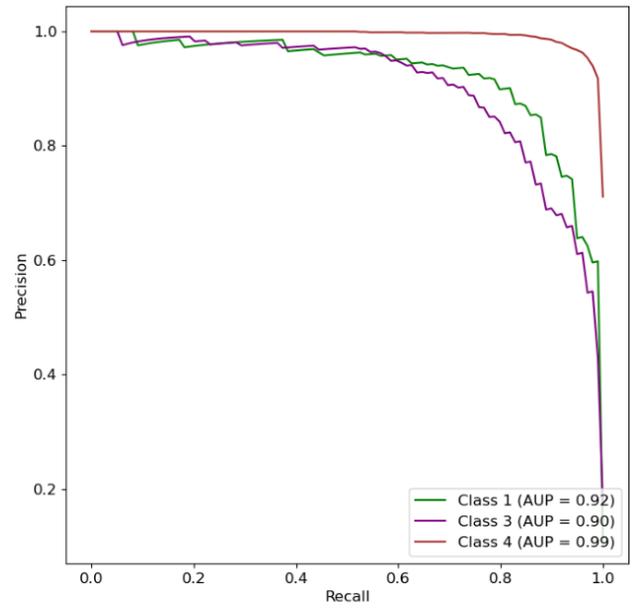

**Figure S18: ROC and Precision-Recall curves for 102 nutrients.** The XGB Classifier attained AUC scores of 0.99, 0.97, and 0.98 for NOVA Classes 1, 3, and 4, respectively. Additionally, it achieved AUP scores of 0.92, 0.90, and 0.99 for NOVA Classes 1, 3, and 4, respectively.

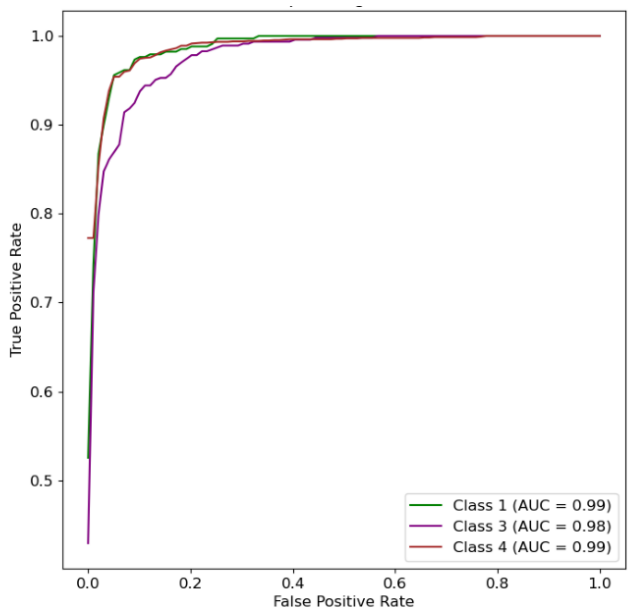 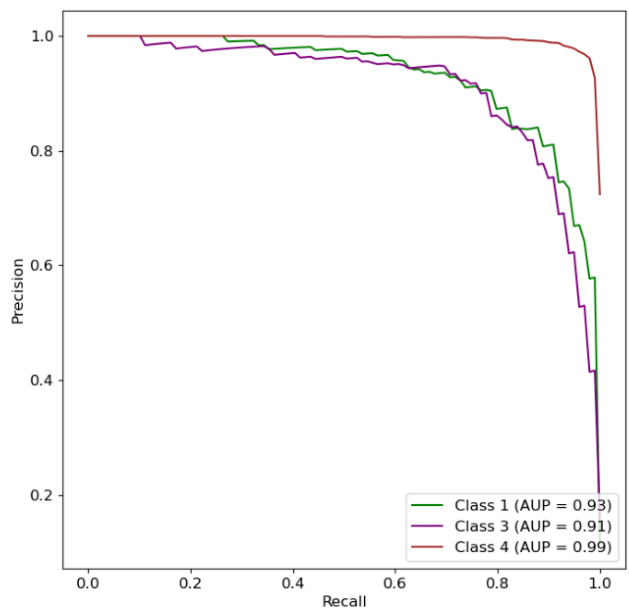

**Figure S19: ROC and Precision-Recall curves for 65 nutrients.** The Gradient Boost Classifier attained AUC scores of 0.99, 0.96, and 0.99 for NOVA Classes 1, 3, and 4, respectively. Additionally, it achieved AUP scores of 0.93, 0.91, and 0.99 for NOVA Classes 1, 3, and 4, respectively.

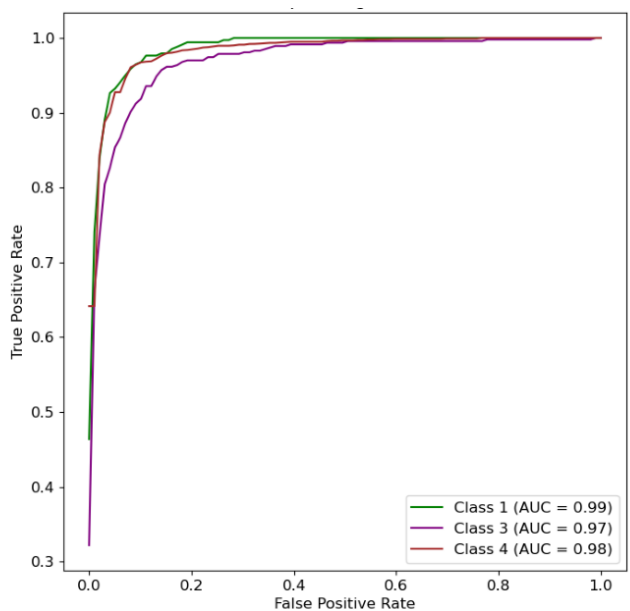 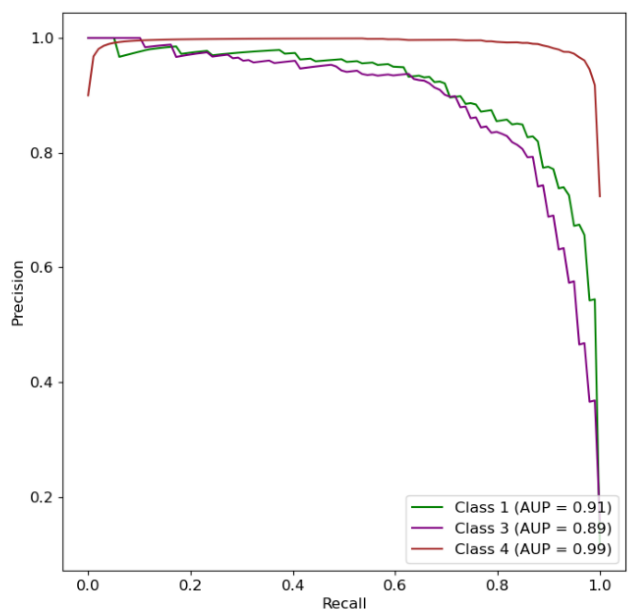

**Figure S20: ROC and Precision-Recall curves for 13 nutrients.** The Gradient Boost Classifier attained AUC scores of 0.99, 0.97, and 0.98 for NOVA Classes 1, 3, and 4, respectively. Additionally, it achieved AUP scores of 0.91, 0.89, and 0.99 for NOVA Classes 1, 3, and 4, respectively.

For the 102-nutrient panel, the merger of NOVA Class 2 yielded an F1-score of 0.9357 and MCC of 0.8558 for the best-performing model, i.e. Gradient Boost. In contrast, removing NOVA Class 2 yielded an F1-score of 0.9387 and MCC of 0.8626 for the best-performing model, i.e. XGB. Compared to the experiment with all the classes, the best-performing model, i.e. LGBM, achieved an F1-score of 0.9411 and MCC of 0.8691. Through this comparison, we can conclude that the absence of or the merger of NOVA Class 2 doesn't affect the model performance significantly for the 102-nutrient panel.

For the 65-nutrient panel, the merger of NOVA Class 2 yielded an F1-score of 0.9328 and MCC of 0.8499 for the best-performing model, i.e. Random Forest. In contrast, removing NOVA Class 2 yielded an F1-score of 0.9352 and MCC of 0.8521 for the best-performing model, i.e. Gradient Boost. Compared to the experiment with all the classes, the best-performing model, i.e. Random Forest, achieved an F1-score of 0.9388 and MCC of 0.8646. Through this comparison, we can conclude that the absence of or the merger of NOVA Class 2 doesn't affect the model performance significantly for the 65-nutrient panel.

For the 13-nutrient panel, the merger of NOVA Class 2 yielded an F1-score of 0.9247 and MCC of 0.8331 for the best-performing model, i.e. XGB. In contrast, removing NOVA Class 2 yielded an F1-score of 0.9378 and MCC of 0.8574 for the best-performing model, i.e. Gradient Boost. Compared to the experiment with all the classes, the best-performing model, i.e. Gradient Boost, achieved an F1-score of 0.9284 and MCC of 0.8425. Through this comparison, we can conclude that the absence of or the merger of NOVA Class 2 doesn't affect the model performance significantly for the 13-nutrient panel.

## 11. ROC and Precision Recall Curves for 102 nutrients utilizing pre-trained word embeddings

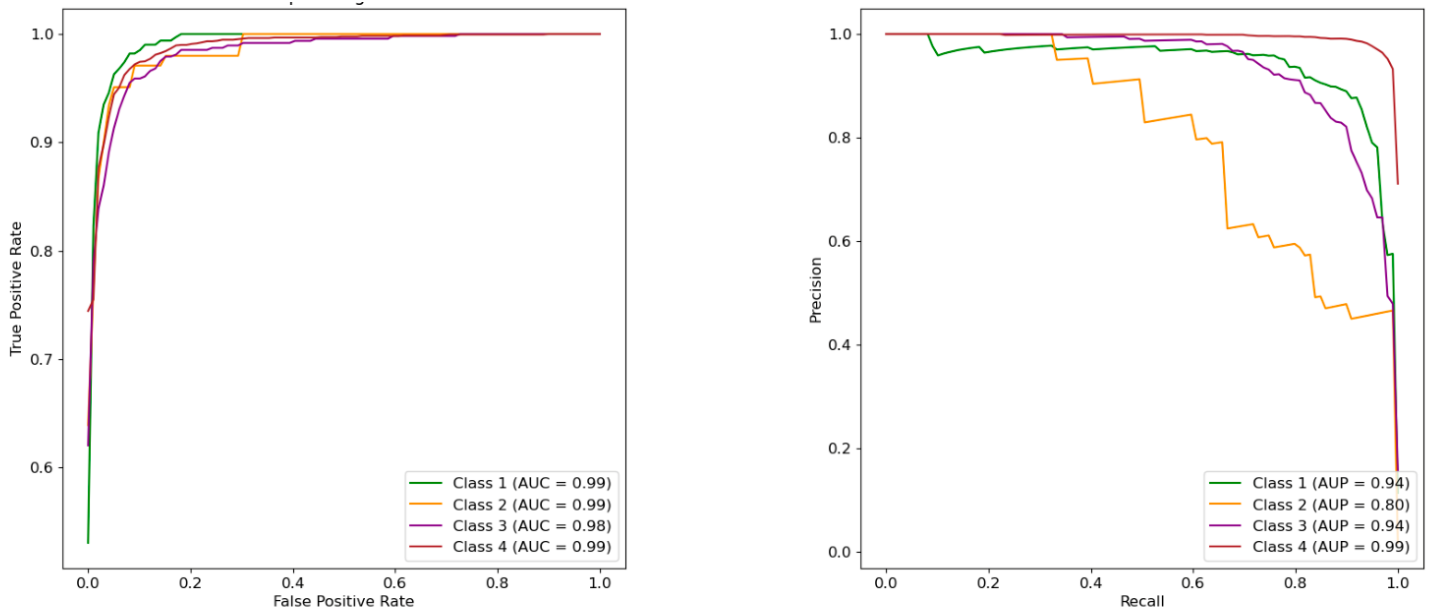

**Figure S21: ROC and Precision-Recall curves for 102 nutrients, and XLM-RoBERTa.** The Gradient Boost Classifier with XLM-RoBERTa as the pre-trained word embedding performed the best, attaining AUC scores of 0.99, 0.99, 0.98, and 0.99 for NOVA Classes 1, 2, 3, and 4, respectively. Additionally, it achieved AUP scores of 0.94, 0.80, 0.94, and 0.99 for NOVA Classes 1, 2, 3, and 4, respectively.

## 12. ROC and Precision Recall Curves for 65 nutrients utilizing pre-trained word embeddings

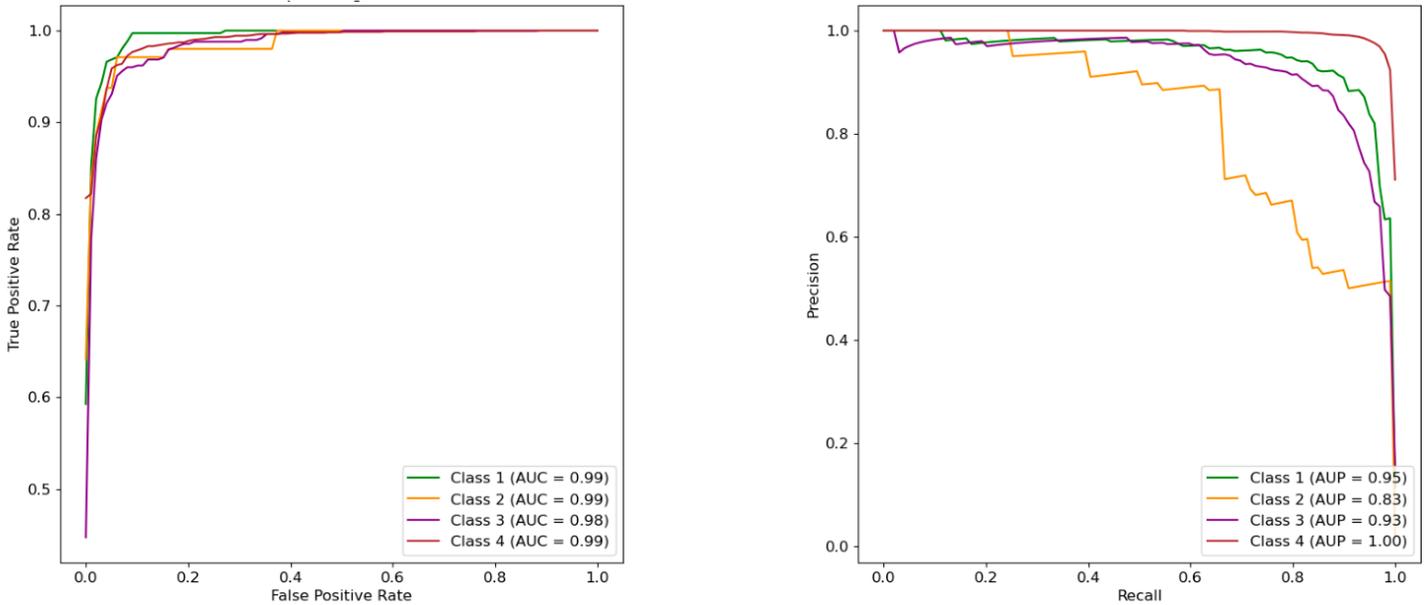

**Figure S22: ROC and Precision-Recall curves for 65 nutrients, and DistilBERT.** The LGBM Classifier with DistilBERT as the pre-trained word embedding performed the best, attaining AUC scores of 0.99, 0.99, 0.98, and 0.99 for NOVA Classes 1, 2, 3, and 4, respectively. Additionally, it achieved AUP scores of 0.95, 0.83, 0.93, and 1.00 for NOVA Classes 1, 2, 3, and 4, respectively.

## 13. ROC and Precision Recall Curves for 13 nutrients utilizing pre-trained word embeddings

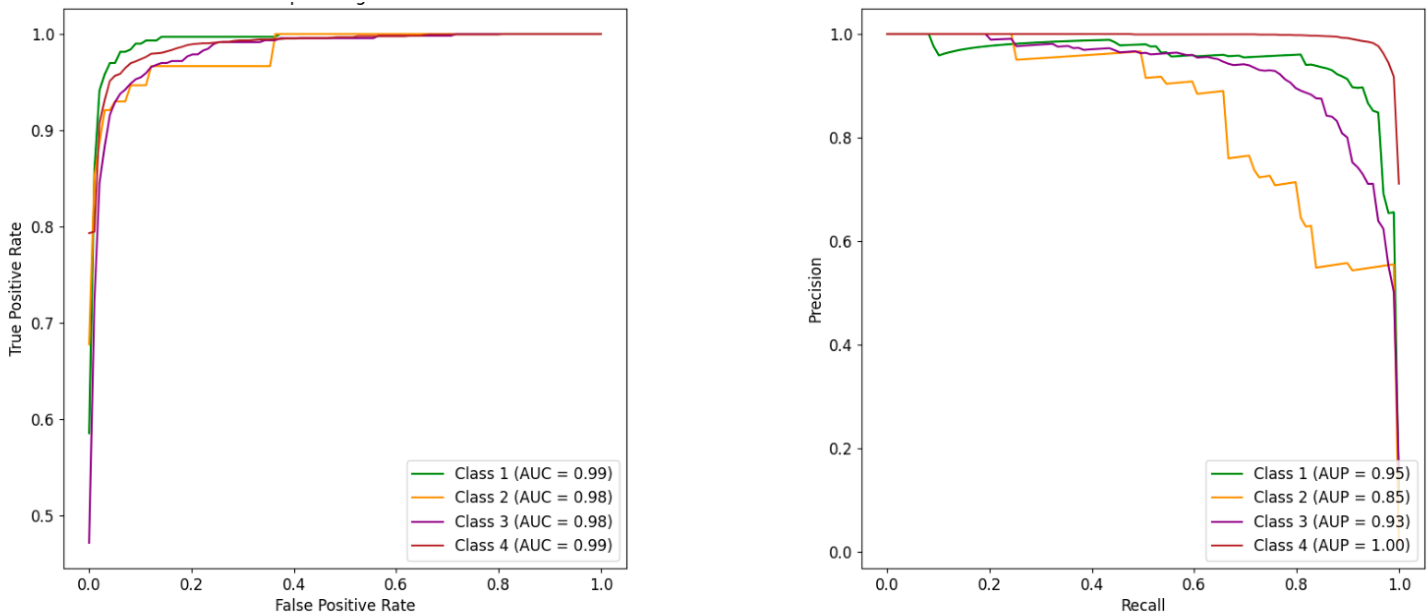

**Figure S23: ROC and Precision-Recall curves for 13 nutrients, and GPT-2.** The LGBM Classifier with GPT-2 as the pre-trained word embedding performed the best, attaining AUC scores of 0.99, 0.98, 0.98, and 0.99 for NOVA Classes 1, 2, 3, and 4, respectively. Additionally, it achieved AUP scores of 0.95, 0.85, 0.93, and 1.00 for NOVA Classes 1, 2, 3, and 4, respectively.

## Additional Supplementary Files

[Supplementary Data 2](#)

[Supplementary Data 3](#)

[Supplementary Data 4](#)

[Supplementary Data 5](#)

[Supplementary Data 6](#)

[Supplementary Data 7](#)